\documentclass[11pt, a4paper]{article}

\usepackage{fullpage}
\usepackage{latexsym}
\usepackage{psfig}
\usepackage{amsfonts}
\usepackage{amssymb}

\newcommand{\beq}{\begin{equation}}
\newcommand{\eeq}{\end{equation}}
\newcommand{\ba}{\begin{array}}
\newcommand{\ea}{\end{array}}
\newcommand{\bea}{\begin{eqnarray}}
\newcommand{\eea}{\end{eqnarray}}
\newcommand{\bc}{\begin{center}}
\newcommand{\ec}{\end{center}}
\newcommand{\bt}{\begin{table}}
\newcommand{\et}{\end{table}}

\newcommand{\la}[1]{\label{#1}}
\newcommand{\p}{\partial}

\newcommand{\bbox}{\rule{3mm}{3mm}}

\newcommand{\ds}{\displaystyle}
\newcommand{\no}{\noindent}
\newcommand{\pp}[2]{{\partial #1 \over \partial #2}}
\newcommand{\ppn}[3]{{\partial^{#1} #2 \over \partial #3^{#1}}}

\newcommand{\mbf}[1]{\mbox{\boldmath {$#1$}}}

\newcommand{\rf}[1]{(\ref{#1})}
\newcommand{\beqno}{\begin{displaymath}}
\newcommand{\eeqno}{\end{displaymath}}

\newcommand{\been}{\begin{enumerate}}
\newcommand{\een}{\end{enumerate}}

\newlength{\myheight}
\newlength{\mylength}

\newcounter{saveeqn}

\newtheorem{lemma}{Lemma}

\newtheorem{theo}{Theorem}

\newtheorem{example}{Example}

\newcommand{\x}{\mbf{x}}
\newcommand{\y}{\mbf{y}}
\newcommand{\X}{\mbf{X}}
\newcommand{\Y}{\mbf{Y}}
\newcommand{\n}{\mbf{n}}
\newcommand{\approxleq}{\stackrel{<}{\approx}}
\renewcommand{\k}{\mbf{k}}
\newcommand{\T}{\mbf{T}}
\renewcommand{\v}{\mbf{v}}

\newcommand{\C}{{\rm \mbox{C\hspace{-.455em}\rule{0.035em}{0.67em}\hspace{.5em}}}}

\begin{document}

\title{Computing Riemann Theta Functions}

\author{
Bernard Deconinck\thanks{
Department of Mathematics,
Colorado State University,
Fort Collins, CO 80523-1874, USA} 
\and 
Matthias Heil\thanks{
Fachbereich Mathematik,
Technische Universit\"at Berlin,
Strasse des 17.Juni 136,
10623 Berlin, Germany}
\and 
Alexander Bobenko$^\dagger$
\and
Mark van Hoeij\thanks{Department of Mathematics,
Florida State University, Tallahassee, FL 32306, USA}
\and
Marcus Schmies$^\dagger$
~\\~\\
{\small \bf AMS subject classification}:\\
{\small 14K25, 30E10, 33F05, 65D20}~
}

\maketitle

\begin{abstract}

The Riemann theta function is a complex-valued function of $g$ complex
variables. It appears in the construction of many (quasi-) periodic solutions of
various equations of mathematical physics. In this paper, algorithms for its
computation are given. First, a formula is derived allowing the pointwise
approximation of Riemann theta functions, with arbitrary, user-specified
precision. This formula is used to construct a uniform approximation formula,
again with arbitrary precision. 

\end{abstract}

\section{Motivation}

An Abelian function is a $2g$-fold periodic, meromorphic function of $g$
complex variables. Thus, Abelian functions are generalizations of elliptic
functions to more than one variable. Just as elliptic functions are associated
with elliptic surfaces, $i.e.,$ Riemann surfaces of genus 1, Abelian functions
are associated to Riemann surfaces of higher genus. As in the elliptic case,
any Abelian function can be expressed as a ratio of homogeneous polynomials of
an auxiliary function, the Riemann theta function \cite{igusa}. Many
differential equations of mathematical physics have solutions that are written
in terms of Abelian functions, and thus in terms of Riemann theta functions
(see \cite{belokolos, dubrovin} and references therein). Thus, to compute these
solutions, one needs to compute either Abelian functions or theta functions. 
Whittaker and Watson \cite[\S 21.1]{ww} state ``When it is desired
to obtain definite numerical results in problems involving Elliptic functions,
the calculations are most simply performed with the aid of certain auxiliary
functions known as {\em Theta-functions}.'' Whittaker and Watson are referring
to Jacobi's theta functions, but the same can be said for Abelian functions,
using the Riemann theta function. This paper addresses the problem of
computing values of the Riemann theta function and its derivatives. 

\section{Definition}

The Riemann Theta function is defined by 

\beq\la{theta}
\theta(\mbf{z}|\mbf{\Omega})=\sum_{\mbf{n}\in \mathbb{Z}^g} e^{2\pi i
\left(\frac{1}{2}\mbf{n}\cdot \mbf{\Omega} \cdot \mbf{n}+\mbf{n}\cdot \mbf{z}
\right)}, 
\eeq

\no where $\mbf{z}\in \mathbb{C}^g$, $\mbf{\Omega} \in \mathbb{C}^{g\times g}$,
such that $\mbf{\Omega}$ is symmetric ($\mbf{\Omega}^T=\mbf{\Omega}$) and the
imaginary part of $\mbf{\Omega}$, Im$(\mbf{\Omega})$, is strictly positive
definite. Such an $\mbf{\Omega}$ is called a Riemann matrix. Also,
$\mbf{n}\cdot\mbf{z}=\sum_{i=1}^g n_i z_i$, the scalar product of the integer
vector $\mbf{n}$ with $\mbf{z}$; and $\mbf{n}\cdot \mbf{\Omega}\cdot
\mbf{n}=\sum_{i,j=1}^g \Omega_{ij} n_i n_j$. The positive definiteness of
Im$(\mbf{\Omega})$ guarantees the convergence of \rf{theta}, for all values of
$\mbf{z}$. Then the series \rf{theta} converges absolutely in both $\mbf{z}$
and $\mbf{\Omega}$, and uniformly on compact sets. Thus, it defines a
holomorphic function of both $\mbf{z}$ and $\mbf{\Omega}$. The function defined
by \rf{theta} is also known as a multidimensional theta function, or as a theta
function of several variables. There are as many different conventions for
writing the Riemann theta function as there are names for it. These different
conventions differ from \rf{theta} by at worst a complex scaling transformation
on the arguments. An extensive overview of the wealth of properties
of $\theta(\mbf{z}|\mbf{\Omega})$ is found in \cite{tata1, tata2, tata3}. 

\subsection*{Remarks}

\begin{itemize}

\item The Riemann theta function was devised by Riemann as a generalization of
Jacobi's theta functions of one variable \cite{Jacobi}, for solving the Jacobi
inversion problem on general compact connected Riemann surfaces
\cite{Riemann}.  For these purposes Riemann considered only theta functions
associated with Riemann surfaces. Riemann theta functions in their full
generality, as defined in \rf{theta}, were considered first by Wirtinger
\cite{wirtinger}, whose convention for the arguments is adopted here. 

\item Often, so-called Riemann theta functions with characteristics are
considered \cite{tata1}. Such theta functions with characteristics are up to an
exponential factor Riemann theta functions \rf{theta} evaluated at a shifted
argument. Thus the computation of the Riemann theta function \rf{theta} also
allows the computation of theta functions with arbitrary characteristics. 

\item In many applications, the Riemann theta function \rf{theta} originates
from a specific Riemann surface, $i.e.$, the Riemann matrix $\mbf{\Omega}$ is
the normalized periodmatrix of the Riemann surface. Obtaining this periodmatrix
is a nontrivial problem, which is now also reduced to a black-box program. This
issue was addressed in \cite{dvh1}. 

\item Some of the algorithms given in this paper have been used already by some
of the authors  (For instance, tori with constant mean curvature, and Willmore
tori with umbilic lines were constructed in \cite{heil} using the results of
\cite{bobenko2, bobenko1}; multiphase solutions of the Kadomtsev-Petviashvili
equation were constructed using Schottky uniformization in \cite{belokolos, 
bobenko3}) but they were not easily available for use by others. Now, these
algorithms are implemented  as black-box programs in Maple and Java. The maple
implementation is included  in the Maple distribution as of Maple 8. It is also
available from {\tt http://www.math.fsu.edu/\verb+~+hoeij/RiemannTheta/}. 
The Java implementation is available from {\tt
www-sfb288.math.tu-berlin.de/\verb+~+jem}. These implementations are discussed in the
appendices. 

\end{itemize}

\section{Rewriting the Riemann theta function} 

The Fourier series representation \rf{theta} is the starting point for the
computation of Riemann theta functions. Separating both $\mbf{z}$ and
$\mbf{\Omega}$ in their real and imaginary parts, $\mbf{z}=\x+i\y$,
$\mbf{\Omega}=\X+i\Y$, and using $\Y=\Y^T$, we obtain 

\bea\nonumber
\theta(\mbf{z}|\mbf{\Omega})&=&\sum_{\n\in \mathbb{Z}^g} e^{2\pi
i\left(\frac{1}{2}\n\cdot \mbf{\Omega}\cdot\n+\n\cdot\mbf{z} \right)}\\
\nonumber
&=&\sum_{\n\in \mathbb{Z}^g} e^{2\pi
i\left(\frac{1}{2}\n\cdot
\X\cdot\n+\n\cdot\x+i\left(\frac{1}{2}\n\cdot\Y\cdot\n+\n\cdot\y\right) 
\right)}\\\nonumber
&=&\sum_{\n\in \mathbb{Z}^g} e^{2\pi
i\left(\frac{1}{2}\n\cdot
\X\cdot\n+\n\cdot\x\right)}e^{-2\pi\left(\frac{1}{2}\n\cdot\Y\cdot\n+
\n\cdot\y\right)}\\\nonumber
&=&\sum_{\n\in \mathbb{Z}^g} e^{2\pi
i\left(\frac{1}{2}\n\cdot
\X\cdot\n+\n\cdot\x\right)}
e^{-2\pi\left(\frac{1}{2}(\n+\mbf{Y}^{-1}\y)\cdot\Y\cdot(\n+\mbf{Y}^{-1}\y)+
-\frac{1}{2}\y\cdot\mbf{Y}^{-1}\cdot\y\right)}\\\la{intermed}
&=&e^{\pi \y\cdot \Y^{-1}\cdot \y}\sum_{\n\in \mathbb{Z}^g} e^{2\pi
i\left(\frac{1}{2}\n\cdot
\X\cdot\n+\n\cdot\x\right)}
e^{-\pi\left(\n+\mbf{Y}^{-1}\y\right)\cdot\Y\cdot\left(\n+\mbf{Y}^{-1}\y
\right)}
\eea

\no At this point, all exponential growth has been isolated: the factor
multiplying the sum grows double-exponentially as the components of $\mbf{z}$
leave the real line, due to the fact that $\Y^{-1}$ is strictly
positive definite: this follows since $\Y$ is strictly positive definite, which
also guarantees the existence of $\Y^{-1}$. All terms remaining in the sum are
either oscillating (the first factor of every term), or a damped exponential
(the second factor of every term). 

Since for most applications, the exponential growth term cancels out, it will be
disregarded in almost all that follows: {\bf any statements of approximation,
pointwise or uniform, of the Riemann theta function pertain to the infinite sum
in \rf{intermed}, without considering the exponential growth}. 

Let $[\mbf{V}]$ be the vector with integer component closest to $\mbf{V}$, and
let $[[\mbf{V}]]=\mbf{V}-[\mbf{V}]$.  Continuing the rewriting of the theta
function, 

\bea\nonumber
\theta(\mbf{z}|\mbf{\Omega})&=&
e^{\pi \y\cdot \Y^{-1} \cdot \y}
\!\!\sum_{\n\in \mathbb{Z}^g} \!e^{2\pi
i\left(\frac{1}{2}\n\cdot
\X\cdot\n+\n\cdot\x\right)}
e^{-\pi\left(\n+\left[\Y^{-1}\y\right]+
\left[\left[\Y^{-1}\y\right]\right]\right)\cdot\Y
\cdot\left(\n+\left[\Y^{-1}\y\right]+\left[\left[\Y^{-1}\y\right]\right]
\right)}\\\nonumber
&=&
e^{\pi \y\cdot \Y^{-1} \cdot \y}
\sum_{\n\in \mathbb{Z}^g} e^{2\pi
i\left(\frac{1}{2}\left(\n-\left[\Y^{-1}\y\right]\right)\cdot
\X\cdot\left(\n-\left[\Y^{-1}\y\right]\right)+
\left(\n-\left[\Y^{-1}\y\right]\right)\cdot\x\right)}\times\\\la{rewrite}
&&~~~~~~~~~~~~~~~~~~\times 
e^{-\pi\left(\n+
\left[\left[\Y^{-1}\y\right]\right]\right)\cdot\Y
\cdot\left(\n+\left[\left[\Y^{-1}\y\right]\right]
\right)}.
\eea

\no The last step is achieved by shifting the summation index $\n$. From this
formulation, it is clear that the size of each one of the terms in the infinite
sum is controlled by its second exponential factor. 

Similarly, formulae are worked out for the derivatives of theta functions.
Denote the $N$-th order directional derivative of the $g$-variable Riemann
theta function along the $g$-dimensional vectors $\k^{(1)}$, $\ldots$,
$\k^{(N)}$ as 

\beq\la{dirders}
D(\k^{(1)}, \ldots, \k^{(N)})\theta(\mbf{z}|\mbf{\Omega})=
\sum_{i_1, \ldots, i_N=1}^g \k^{(1)}_{i_1}\ldots \k^{(N)}_{i_N}
\frac{\p^N\theta(\mbf{z}|\mbf{\Omega})}{\p z_{i_1} \ldots \p z_{i_N}}.
\eeq
Then 
\bea\nonumber
D(\k^{(1)}, \ldots, \k^{(N)})\theta(\mbf{z}|\mbf{\Omega})\!\!\!\!&=&
\!\!\!\!(2\pi i)^N
\sum_{\mbf{n}\in \mathbb{Z}^g}(\k^{(1)}\cdot\n)\ldots(\k^{(N)}\cdot\n) e^{2\pi i
\left(\frac{1}{2}\mbf{n}\cdot \mbf{\Omega} \cdot \mbf{n}+\mbf{n}\cdot \mbf{z}
\right)}\\\nonumber
\!\!\!\!&=&\!\!\!\!(2\pi i)^N e^{\pi \y\cdot \Y^{-1}\cdot \y}
\!\!\sum_{\n\in \mathbb{Z}^g}\!\! \left(\k^{(N)}\!\cdot\!\left(\n-
\left[\Y^{-1}\y\right]\right)\right)\!\ldots\! 
\left(\k^{(N)}\!\cdot\!\left(\n-\left[\Y^{-1}\y\right]\right)\right)\!\times\\
\nonumber
&&~~~~~~~~~~~~~~~~\times  e^{2\pi
i\left(\frac{1}{2}\left(\n-\left[\Y^{-1}\y\right]\right)\cdot
\X\cdot\left(\n-\left[\Y^{-1}\y\right]\right)+
\left(\n-\left[\Y^{-1}\y\right]\right)\cdot\x\right)}\!\times\\\la{rewrited}
&&~~~~~~~~~~~~~~~~\times 
e^{-\pi\left(\n+
\left[\left[\Y^{-1}\y\right]\right]\right)\cdot\Y
\cdot\left(\n+\left[\left[\Y^{-1}\y\right]\right]
\right)}, 
\eea

\no using similar steps as before. The exponential growth is still factored
out, but the remaining infinite sum now contains terms that grow algebraically
as the argument of the Riemann theta function leaves the real line. The order
of this growth is equal to the order of the derivative. However, the size of
individual terms within the infinite sum is again determined by the last factor
which is exponentially decaying. 
 
\no {\bf Remark.} Riemann theta functions of genus greater than one have been 
computed before, for instance in \cite{dfs}, where genus three was considered.
There four distinct representations of these Riemann theta functions were used,
based on whether the different phases behaved as trigonometric or hyperbolic
functions (two limits of elliptic functions), depending on the parameters in
the Riemann matrix. These distinct representations were required to have a
manageable number of contributing terms to the series. Such a variety of
representations is not needed here: all limit cases are incorporated in the
form \rf{rewrite}. This is obtained by the separation of the real and imaginary
parts of both the argument $\mbf{z}$ and the Riemann matrix $\mbf{\Omega}$. 

\section{Pointwise Approximation}

The formulae \rf{rewrite} and \rf{rewrited} are the basis for the approximation
of theta functions and their derivatives. All approximations are based on
determining which terms of the infinite sum are dominant. 

For the Riemann theta function, 

\bea\nonumber
\theta(\mbf{z}|\mbf{\Omega}) e^{-\pi \y\cdot \Y^{-1} \cdot \y}&=&
\sum_{\n\in \mathbb{Z}^g} e^{2\pi
i\left(\frac{1}{2}\left(\n-\left[\Y^{-1}\y\right]\right)\cdot
\X\cdot\left(\n-\left[\Y^{-1}\y\right]\right)+
\left(\n-\left[\Y^{-1}\y\right]\right)\cdot\x\right)}\times\\\nonumber
&&~~~~~~~\times 
e^{-\pi\left(\n+
\left[\left[\Y^{-1}\y\right]\right]\right)\cdot\Y
\cdot\left(\n+\left[\left[\Y^{-1}\y\right]\right]
\right)}\\\nonumber
&=&\lim_{R\rightarrow \infty} \sum_{S_R} e^{2\pi
i\left(\frac{1}{2}\left(\n-\left[\Y^{-1}\y\right]\right)\cdot
\X\cdot\left(\n-\left[\Y^{-1}\y\right]\right)+
\left(\n-\left[\Y^{-1}\y\right]\right)\cdot\x\right)}\times\\
&&~~~~~~~~~~~\times 
e^{-\pi\left(\n+
\left[\left[\Y^{-1}\y\right]\right]\right)\cdot\Y
\cdot\left(\n+\left[\left[\Y^{-1}\y\right]\right]
\right)},
\eea

\no where 
\beq
S_R=\left\{\n \in \mathbb{Z}^g | \pi\left(\n+
\left[\left[\Y^{-1}\y\right]\right]\right)\cdot\Y
\cdot\left(\n+\left[\left[\Y^{-1}\y\right]\right]
\right)<R^2\right\}.
\eeq 

\no Below, we show this limit exists, hence proving that the Riemann theta
function is well defined. This proof is different from that found in many
references (see \cite{tata1}, for example) in that it also allows the
estimation of the error made in the approximation of the truncation of the
series by only considering a finite value of $R$.  

Since $\Y$ is strictly positive definite, it has a Cholesky decomposition,
$\Y=\T^T \T$. Let $\Lambda$ be the lattice of all vectors $\v(\n)$ of the form
$\v(\n)=\sqrt{\pi}\,\T \left(\n+\left[\left[\Y^{-1}\y\right]\right]\right)$,
for $\n \in \mathbb{Z}^g$. Then  $S_R$ is rewritten as 

\beq\la{SR}
S_R=\left\{\v(\n) \in \Lambda \Big| ||\v(\n)||<R\right\}.
\eeq

\no Thus, the approximation
requires finding all lattice points in $\Lambda$ that are also inside the
$g$-dimensional sphere $||v(n)||=R$.

\no Let 
\bea
\theta_R(\mbf{z}|\mbf{\Omega})&=&e^{\pi \y\cdot \Y^{-1} \cdot \y}
\sum_{S_R} e^{2\pi
i\left(\frac{1}{2}\left(\n-\left[\Y^{-1}\y\right]\right)\cdot
\X\cdot\left(\n-\left[\Y^{-1}\y\right]\right)+
\left(\n-\left[\Y^{-1}\y\right]\right)\cdot\x\right)} e^{-||\v(\n)||^2},
\eea

\no then 
\bea
\epsilon(R)&=&
\left|\theta(\mbf{z}|\mbf{\Omega})-\theta_R(\mbf{z}|\mbf{\Omega})\right|
 e^{-\pi \y\cdot \Y^{-1} \cdot \y}\\\nonumber
 &=&
\left|\sum_{\Lambda\setminus S_R} e^{2\pi
i\left(\frac{1}{2}\left(\n-\left[\Y^{-1}\y\right]\right)\cdot
\X\cdot\left(\n-\left[\Y^{-1}\y\right]\right)+
\left(\n-\left[\Y^{-1}\y\right]\right)\cdot\x\right)}e^{-||\v(\n)||^2}\right|\\
\nonumber
&\leq&
\sum_{\Lambda\setminus S_R} \left|e^{2\pi
i\left(\frac{1}{2}\left(\n-\left[\Y^{-1}\y\right]\right)\cdot
\X\cdot\left(\n-\left[\Y^{-1}\y\right]\right)+
\left(\n-\left[\Y^{-1}\y\right]\right)\cdot\x\right)}e^{-||\v(\n)||^2}\right|\\
\la{firsterror}
&=&\sum_{\Lambda\setminus S_R} e^{-||\v(\n)||^2}.
\eea

Thus $\epsilon(R)$ is the absolute error of the oscillatory part of the Riemann
theta function, due to the truncation to a finite number of terms. In practice,
this oscillatory part is of order 1, so $\epsilon(R)$ is also a measure of the
relative error. 
 
In order to estimate this last sum, the following theorems and lemmas are used. 

\begin{theo}[Mean-value Theorem for subharmonic functions]\la{theo:1}
Let $\Omega$ be a
domain in $\mathbb{R}^g$. Let $u$  be a twice
continuously differentiable function on $\Omega$, continuous on the boundary of
$\p\Omega$ of $\Omega$, which is subharmonic on $\Omega$. 
Then for any ball $B_R^g(\y)$ in $\Omega$ of 
radius $R$ and center $\y$
\beq\nonumber
\ds u(\y)\leq \frac{1}{R^g {\rm Vol} B_1^g(0)}\int_{B_R^g(\y)}u(\x) d\x.  
\eeq
\end{theo}

\no A proof of this theorem is found in \cite{gilbarg}.

\begin{lemma}\la{lemma:1}
For every integer $p\geq 0$ and any dimension $g>0$ 
the function $u(\y)=e^{-||\y||^2}||\y||^p$ with $\y \in \mathbb{R}^g$ 
is subharmonic on
$\mathbb{R}^g\setminus B^g_R(0)$, with
$R=\frac{1}{2}\sqrt{g+2p+\sqrt{g^2+8p}}$. 
\end{lemma}

\no{\bf Proof.} Since $u(\y)$ depends only on $r=||\y||$, this calculation is
done in $g$-dimensional spherical coordinates. 
\bea\nonumber
\Delta u(\y)&=&\ppn{2}{u}{r}+\frac{g-1}{r}\pp{u}{r}\\\nonumber
&=&2 u(\y)\left(2r^2-(g+2p)+\frac{p^2+(g-2)p}{2r^2}\right)
\eea
This last factor is positive if $r>\frac{1}{2}\sqrt{g+2p+\sqrt{g^2+8p}}$, from
which the result follows. \hfill $\bbox$

\begin{lemma}\la{lemma:2}
Let $\Lambda$ be a $g$-dimensional affine lattice in
$\mathbb{R}^g$, $i.e.$, $\Lambda=\left\{\X\n+\x | \n\in \mathbb{Z}^g\right\}$,
with $\X \in Gl(n,\mathbb{R})$, $\x\in \mathbb{R}^g$, 
and let $p \in \mathbb{Z}$,
positive. Then
\beq\la{lemma}
\sum_{\y\in \Lambda, ||\y||\geq R} ||\y||^pe^{-||\y||^2}
\leq \frac{g}{2}\left(\frac{2}{\rho}\right)^g 
\Gamma\left(\frac{g+p}{2},(R-\rho/2)^2\right),
\eeq
where $\Gamma(z,d)$ $=$ $\int_d^\infty t^{z-1} e^{-z} dz$, the 
incomplete Gamma
function \cite{abst}, $R > \frac{1}{2}\sqrt{g+2p+\sqrt{g^2+8p}}+\rho/2$, 
and $\rho = {\rm min}\left\{||\x-\y||~|\x,\y \in \Lambda,
\x \neq \y \right\}$.
\end{lemma}

\no{\bf Proof.} By the previous lemmas, 
\bea\nonumber
\sum_{\y\in \Lambda, ||\y||\geq R} ||\y||^pe^{-||\y||^2}&\leq&
\sum_{\y\in \Lambda, ||\y||\geq R}  \frac{1}{(\rho/2)^g \mbox{Vol}B_1^g(0)}
\int_{B_{\rho/2}^g(\y)}||\x||^pe^{-||\x||^2}d\x\\\la{gross}
&=&\frac{1}{\mbox{Vol}B_1^g(0)}\left(\frac{2}{\rho}\right)^g
\sum_{\y\in \Lambda, ||\y||\geq R}
\int_{B_{\rho/2}^g(\y)}||\x||^pe^{-||\x||^2}d\x\\\nonumber
&\leq&\frac{1}{\mbox{Vol}B_1^g(0)}\left(\frac{2}{\rho}\right)^g
\int_{\mathbb{R}^g\setminus B^g_{R-\rho/2}(0)}
||\x||^pe^{-||\x||^2}d\x\\\nonumber
&=&\frac{\mbox{Vol}S_1^{g-1}(0)}{\mbox{Vol}B_1^g(0)}
\left(\frac{2}{\rho}\right)^g
\int_{R-\rho/2}^\infty e^{-r^2}r^{g-1+p}dr\\\nonumber
&=&\frac{g}{2} 
\left(\frac{2}{\rho}\right)^g\int_{(R-\rho/2)^2}^\infty e^{-t} t^{(g+p)/2-1}dt\\
\nonumber
&=&\frac{g}{2} 
\left(\frac{2}{\rho}\right)^g\Gamma\left(\frac{g+p}{2},(R-\rho/2)^2\right).
\eea
\no Here Vol$S_r^{g-1}(0)/$Vol$V_r^g(0)=g/r$ was used.\hfill $\bbox$

\vspace*{0.2in}

\begin{figure}[htb]
\centerline{\psfig{file=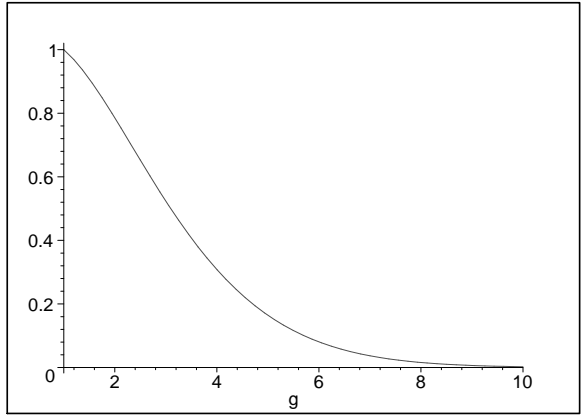,width=3in,angle=270}} 
\caption{\label{fig:fillfactor}
Fill factor $c(\Lambda,\rho/2)$ for the case of a square lattice.}
\end{figure}

{\bf Remark.} The inequality following \rf{gross} is quite crude. Specifically,
it adds all the space in between the balls around the lattice points. 
Taking this in to account, another estimate is as follows:
\bea\nonumber
\sum_{\y\in \Lambda, ||\y||\geq R} ||\y||^pe^{-||\y||^2}&\leq&
\frac{1}{\mbox{Vol}B_1^g(0)}\left(\frac{2}{\rho}\right)^g
\sum_{\y\in \Lambda, ||\y||\geq R}
\int_{B_{\rho/2}^g(\y)}||\x||^pe^{-||\x||^2}d\x\\\nonumber
&\approx&\frac{1}{\mbox{Vol}B_1^g(0)}\left(\frac{2}{\rho}\right)^g 
c(\Lambda,\rho/2)
\int_{\mathbb{R}^g\setminus B^g_{R-\rho/2}(0)}
||\x||^pe^{-||\x||^2}d\x.
\eea
Here $c(\Lambda,\rho/2)$ is the ``fill factor'' of the lattice $\Lambda$ with
balls of radius $\rho/2$: it is the fraction of the lattice volume that is filled
if a ball of radius $\rho/2$ is placed at every lattice point. This estimate is 
justified, since the function being integrated only depends on the radial
variable. Thus, within thin radial shells the function is almost constant,
independent of the proximity to a lattice point. 
The crudeness of the inequality following \rf{gross} is illustrated in Fig.
\ref{fig:fillfactor}, where the fill factor $c(\Lambda,\rho/2)$ is shown for the
scenario of minimal 
$|\Lambda|=\rho^g$, $i.e.,$ a square lattice. Then
$c(\Lambda,\rho/2)=\pi^{g/2}/(2^{g-1}g\Gamma(g/2))$, which decays fast as
$g\rightarrow \infty$. Figure \ref{fig:fillfactor} shows that for a square
lattice, not using the fill factor results in the error being overestimated by
only one digit for $g=6$. Using the fill factor for generic lattices, where
$\Lambda>\rho^g$, usually leads to a more significant improvement. 

\no Thus 
$\ds c(\Lambda,\rho/2)=\mbox{Vol}B^g_{\rho/2}(0)/|\Lambda|=
\frac{2\pi^{g/2}}{g\Gamma(g/2)}\frac{(\rho/2)^g}{|\Lambda|}$, where $|\Lambda|$
is the determinant of the lattice. It is the volume of a cell of 
$\Lambda$ spanned by a set of generating vectors. This gives 
\bea\nonumber
\sum_{\y\in \Lambda, ||\y||\geq R} ||\y||^pe^{-||\y||^2}
&\approxleq&\frac{\mbox{Vol}S_1^{g-1}(0)}{\mbox{Vol}B_1^g(0)}
\left(\frac{2}{\rho}\right)^g
\frac{2\pi^{g/2}}{g\Gamma(g/2)}\frac{(\rho/2)^g}{|\Lambda|}
\int_{R-\rho/2}^\infty e^{-r^2}r^{g-1+p}dr\\\nonumber
&=&g
\frac{2\pi^{g/2}}{g\Gamma(g/2)}\frac{1}{|\Lambda|}
\frac{1}{2}\int_{(R-\rho/2)^2}^\infty e^{-t} t^{(g+p)/2-1}dt\\\la{soft}
&=&\frac{\pi^{g/2}}{|\Lambda|\Gamma(g/2)}
\Gamma\left(\frac{g+p}{2},(R-\rho/2)^2\right).
\eea

Using \rf{lemma} with $p=0$, \rf{firsterror} becomes 
\beq\la{hundredpercent}
\epsilon(R)\leq \sum_{\Lambda\setminus S_R} e^{-||\v(\n)||^2}
\leq
\frac{g}{2}\left(\frac{2}{\rho}\right)^g\Gamma\left(\frac{g}{2},(R-\rho/2)^2
\right),
\eeq
where $\rho$ is the length of the shortest lattice vector in $\Lambda$:
$\rho=\mbox{min}\left\{||x||, x \in \Lambda\right\}$, and
$R>(\sqrt{2g}+\rho)/2$ (to satisfy the subharmonicity condition of Lemma
\ref{lemma:1}). As claimed, this proves that the Riemann theta function
is well-defined, since $\lim_{R\rightarrow\infty}\Gamma(z,R)=0$ \cite{abst}. 

Thus, in order to approximate the oscillatory part of a Riemann 
theta function of $g$ complex variables with a
pre-determined error $\epsilon$, one solves the equation
\beq\la{hundredpercent2}
\epsilon=
\frac{g}{2}\left(\frac{2}{\rho}\right)^g\Gamma\left(\frac{g}{2},(R-\rho/2)^2
\right),
\eeq
for $R$, with $R>(\sqrt{2g}+\rho)/2$, real. If no solution exists, then 
$R=(\sqrt{2g}+\rho)/2$ suffices. These results combine to give 

\begin{theo}[Pointwise Approximation]\la{theo:2}
The Riemann theta function is approximated by 
\beq
\theta(\mbf{z}|\mbf{\Omega})\approx e^{\pi \y\cdot \Y^{-1} \cdot \y}
\sum_{S_R} e^{2\pi
i\left(\frac{1}{2}\left(\n-\left[\Y^{-1}\y\right]\right)\cdot
\X\cdot\left(\n-\left[\Y^{-1}\y\right]\right)+
\left(\n-\left[\Y^{-1}\y\right]\right)\cdot\x\right)} e^{-||\v(\n)||^2},
\eeq
\sloppypar \no 
with absolute error $\epsilon$ on the sum. Here $S_R=\left\{\v(\n)\in
\Lambda\big|~||\v(n)||<R\right\}$,
$\Lambda$ $=$ 
$\left\{\sqrt{\pi}\T(\n+[[\Y^{-1}\y]])~|\n\in \mathbb{Z}^g\right\}$. The shortest
distance between any two points of $\Lambda$ is denoted by $\rho$. Then the 
radius
$R$ is determined as the greater of $(\sqrt{2g}+\rho)/2$ and the real positive
solution of $\epsilon=
g\,2^{g-1}\Gamma\left(g/2,(R-\rho/2)^2\right)/\rho^g$.  
\end{theo}

This theorem gives a pointwise approximation to the Riemann theta function: for
every $\mbf{z}$ at which the Riemann theta function is approximated, $S_R$ is
different, although $R$ is unchanged as long as $g$ and $\epsilon$ remain the
same. The error used in Theorem \ref{theo:2} is referred to as the 100\% Error
(100\%E). 

Another good estimate for $R$ is obtained from \rf{soft}. Then 
$R$ is the greater of $(\sqrt{2g}+\rho)/2$ and the real positive
solution of $\epsilon=
\pi^{g/2}\Gamma\left(g/2,(R-\rho/2)^2\right)/(|\Lambda|\Gamma(g/2))$. This error
is referred to as the Fill Factor Error (FFE). 





\begin{figure}[htb]
\begin{tabular}{cc}
\psfig{file=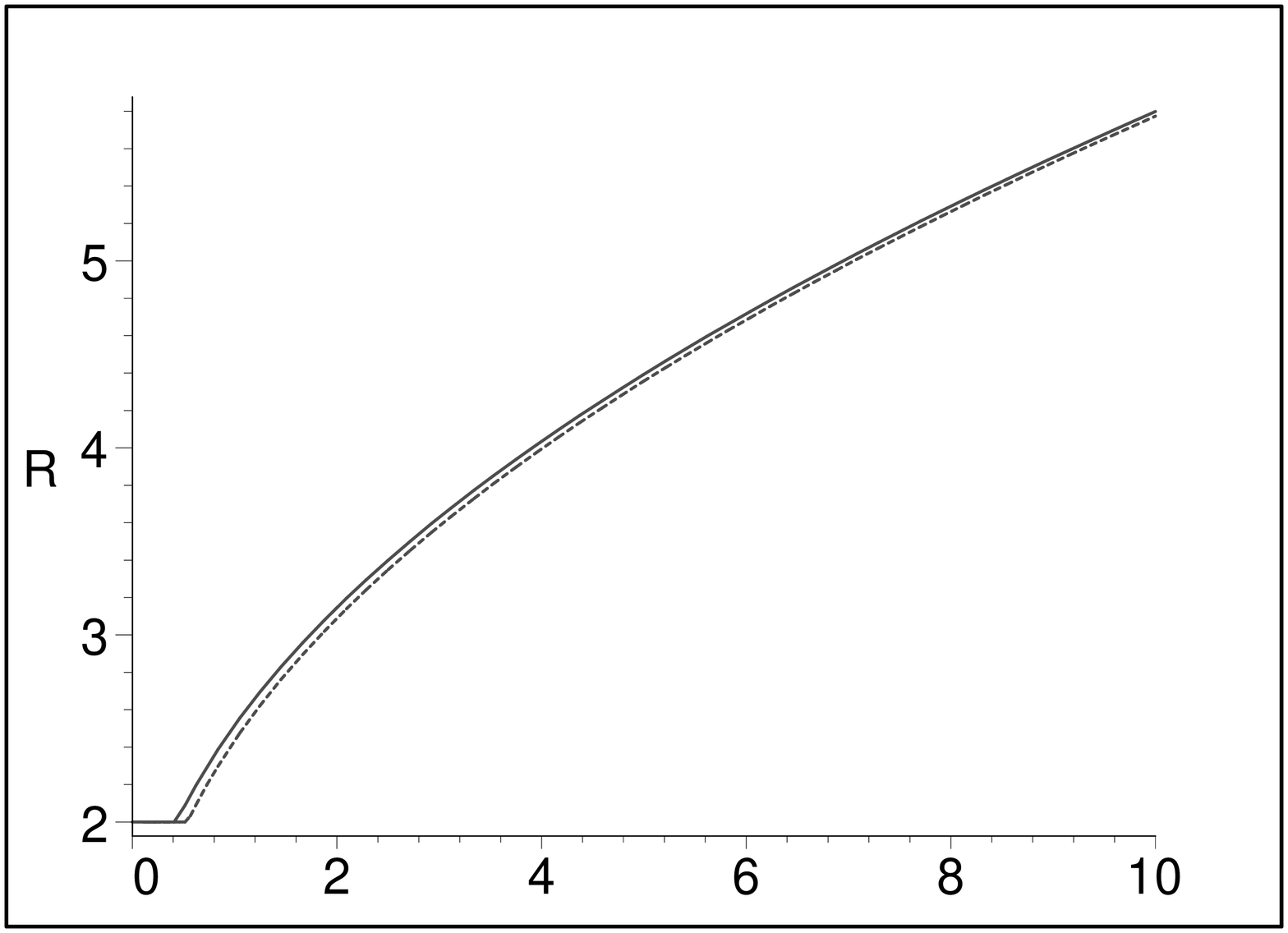,width=3in,angle=270}&
\psfig{file=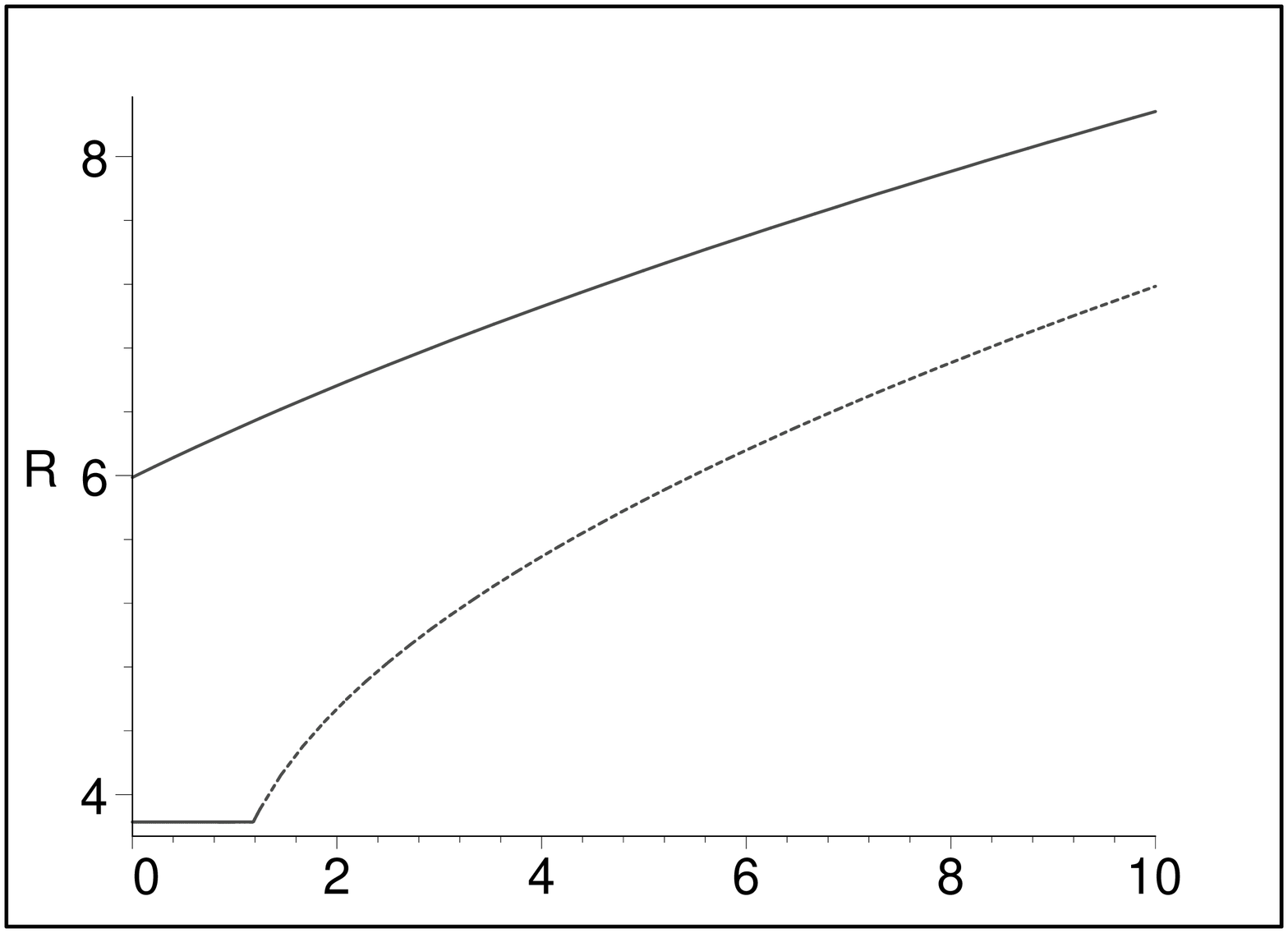,width=3in,angle=270}\\
(a) $g=2$ & (b) $g=16$
\end{tabular}
\caption{\label{fig:errors}
R, assuming the 100\% Error (solid line) and 
Fill Factor Error (dashed line) as functions of
$-log_{10}(\epsilon)$, which is the number of accurate digits.}
\end{figure}

Figure \ref{fig:errors} compares the size $R$ of the ellipsoid, required to
obtain a prescribed error $\epsilon$, as a result of using the 100\%E or the
FFE, for the cases $g=2$ and $g=16$. In these figures, the worst-case scenario 
$|\Lambda|=\rho^g$ was
assumed. As expected, the difference between the 100\%E and the FFE is small
for small genus, but becomes significant for larger genus. 

Two examples are shown in Table \ref{tab:pointwise}, for various values of
$\epsilon$. Both of these use the Riemann matrix $\mbf{\Omega}_g$ with
$\Omega_{jj}=i$, $j=1,\ldots, g$, and $\Omega_{jk}=-1/2$, $j\neq k$. The first
example (left side of the table) corresponds to $g=2$, the second (right side)
to $g=6$. For both examples, the Riemann theta function is evaluated at
$\mbf{z}=\mbf{0}$, thus in this case the oscillatory part of the Riemann theta
function equals the Riemann theta function. The value of the
Riemann theta function is given in the first line. The first column of an
example gives $\epsilon$, used for the pointwise approximation, with either the
100\%E, or the FFE. In the table $N(\mbox{100\%E})$ and $N(\mbox{FFE})$ denote
the number of terms that are used to compute the oscillatory part of the
Riemann theta function, $i.e.$, the number of elements of $S_R$, using the
100\%E or the FFE, respectively. AE denotes Actual Error, $i.e.$, the
difference between the actual value of the oscillatory part of the Riemann
theta function (computed using 30 digits of accuracy, and $\epsilon=10^{-30}$)
and the computed value, using the two different error formulae. 

Using \rf{rewrite}, we see that the first example, with $g=2$, computes
$\sum_{n_1,n_2=-\infty}^{\infty}\cos(n_1n_2\pi)e^{-\pi(n_1^2+n_2^2)}$. Since for
this example $g=2$, there should be little difference between the computation
using the 100\%E or the FFE. This is confirmed: although the value of $R$ is
slightly different for both computations, the number of elements of $S_R$ is
identical, resulting in both computations having the same accuracy. 

For the second example, with $g=6$, there is a difference between the two
computations, although it is not as outspoken as for $g=16$ in Fig.
\ref{fig:errors}. It is clear that the Actual Error using the FFE computation
is closer to $\epsilon$, than using the 100\%E. Even in this case, the
computation does several orders of magnitude better than prescribed. This, of
course, is
due to the inequalities that were used to obtain these error
estimates.    

\begin{table}[htb]
\begin{center}
\begin{tabular}{|l|c|c||l|c|c|c|c|}
\hline
\multicolumn{3}{|c||}{$g=2$}&\multicolumn{5}{|c|}{$g=6$}\\
\multicolumn{3}{|c||}{$\theta(0,0|\mbf{\Omega_2})\,
e^{-\pi\y\cdot\Y_2^{-1}\y}
=1.1654010572$}
&\multicolumn{5}{|c|}{$\theta(0,0,0,0,0,0|\mbf{\Omega_6})\,e^{-\pi
\y\cdot\Y_6^{-1}\y}=8.3721839831$}\\
\hline
\multicolumn{1}{|c|}{$\epsilon$} & $N(\mbox{100\%E})$ & AE(100\%FE)& 
\multicolumn{1}{|c|}{$\epsilon$} & $N(\mbox{100\%E})$ & AE(100\%FE) & $N(\mbox{FFE})$ & AE(FFE)\\
&$=N(\mbox{FFE})$&=AE(FFE)&&&&&\\
\hline 
1E-1&5&7.5E-3& 1E-1& 485& 4.3E-5& 233& 9.2E-4\\
1E-2&9&1.5E-5& 1E-2& 797& 3.8E-6& 485& 4.3E-5\\
1E-3&13&1.2E-6&1E-3& 1341& 2.6E-7& 797& 3.8E-6\\
1E-4&21&5.1E-11&1E-4& 2301& 1.3E-8& 1341& 2.6E-7\\
1E-5&21&5.1E-11&1E-5& 3321& 4.2E-10& 2301& 1.3E-8\\
1E-6&21&5.1E-11&1E-6& 4197& 3.8E-11& 3321& 4.2E-10\\
1E-7&21&5.1E-11&1E-7& 5757& 2.3E-12& 4197& 3.8E-11\\
1E-8&25&1.9E-12&1E-8& 8157& 9.2E-14& 5757& 2.3E-12\\
1E-9&29&1.8E-13&1E-9& 10237& 3.5E-15& 8157& 9.2E-14\\
1E-10&37&1.5E-17&1E-10& 12277& 2.7E-16& 10237& 3.5E-15\\
\hline
\end{tabular}
\end{center}
\vspace*{-0.2in}
\caption{\la{tab:pointwise} Two examples of using the pointwise approximation to
compute the oscillatory part of a Riemann theta function.}
\end{table}

\subsection*{Remarks}

\begin{itemize}

\item The method for approximating the oscillatory part of a Riemann theta
function requires the determination of the elements of $S_R$. This is the set
of all elements of $\Lambda$ that lie inside the $g$-dimensional sphere
determined by $||v(n)||=R$. It is easier to consider the equivalent problem of
determining the points with integer coordinates that lie inside the
$g$-dimensional ellipsoid determined by $\pi (\n-\mbf{c})\cdot \Y \cdot
(\n-\mbf{c})=R^2$, where $c$ is the center of the ellipsoid. This is done
recursively, by remarking that every $(g-1)$-dimensional plane section of a
$g$-dimensional ellipsoid is a $(g-1)$-dimensional sphere. A finite number (due
to the discrete nature of the lattice) of such $(g-1)$-dimensional sections are
taken. All of these are $(g-1)$-dimensional ellipsoids, on which this section
process is repeated, until one arrives at a one-dimensional ellipsoid, $i.e.$,
a line piece. At this level it is easy to determine which integer points are in
this piece. This method takes full advantage of the triangular form of the
Cholesky decomposition of $\Y=\T^T\T$. This is done as follows: all integer points
satisfying 
\beq\la{ellipsoid}
\|\T(\n-\mbf{c})\|<R/\sqrt{\pi}
\eeq
are sought for. Since $\T$ is upper triangular, this implies
$|T_{gg}(n_g-c_g)|<R/\sqrt{\pi}$, or 
$$
c_g-\frac{R}{\sqrt{\pi}T_{gg}}<n_g<c_g+\frac{R}{\sqrt{\pi}T_{gg}}.
$$
This gives a set of allowed values of $n_g$. For each one of these, write
$\n=(\hat{\n},n_g)^T$, $\mbf{c}=(\hat{\mbf{c}},c_g)^T$. With
$$
\mbf{T}=\left(\ba{cc}\hat{\mbf{T}}& \hat{\mbf{t}}\\\mbf{0}& T_{gg} \ea\right),
$$
\rf{ellipsoid} becomes $ \| \hat{\T}(\hat{\n}-\hat{\mbf{c}})+
\hat{\mbf{t}}(n_g-c_g) \|^2+T_{gg}^2
(n_g-c_g)^2 <R^2/\pi$, which
is rewritten as 
$$
\|\hat{\T}(\hat{\n}-\hat{\mbf{c}}+
\hat{\T}^{-1}\hat{\mbf{t}}(n_g-c_g))\|<\sqrt{R^2/\pi-T_{gg}^2 (n_g-c_g)^2}
$$
This is a similar problem as \rf{ellipsoid}, but in dimension $g-1$ instead of
$g$, and with a different $R$, $\mbf{c}$ and $\T$. The procedure is now
repeated. 
\item Both error estimates require the knowledge of $\rho$, the shortest
lattice vector of $\Lambda$. The problem of finding the shortest lattice vector
of a given lattice is one of the main problems in the study of lattices. It is
usually addressed using lattice reduction. Lattice reduction attempts to find a
standard form for the matrix of generating vectors $\mbf{t}_j$, $j=1,\ldots, g$
of the lattice, in which certain intrinsic ($i.e.$, independent of the
representation of the generating vectors; such as $\rho$, for instance)
properties of the lattice are easier to examine. The problem of computing
$\rho$ is known to be $NP$ hard in the parameters $g$ and
$\ln(\mbox{max}_j(||t_j||))$ \cite{vallee}. However, an approximate algorithm
due to Lenstra, Lenstra and Lov\'asz (the LLL algorithm, \cite{lll}) is known
which is polynomially hard in $g$ and 
$\ln(\mbox{max}_j(||t_j||))$. This approximate
algorithm is exact in low dimensions (guaranteed in 1,2 and 3), 
but only gives approximate answers in
high dimensions. In practice, the algorithm has been found very satisfactory. 
A rigorous error estimation is possible using the results of the LLL algorithm,
as it provides both a lower and an upper bound for $\rho$: If $\hat{\rho}$ is
the value of $\rho$ found by the LLL algorithm, then 
$\hat{\rho}/2^{(g-1)/2}\leq\rho\leq\hat{\rho}$. Since the LLL algorithm 
for relatively low genera ($g<10$) usually finds the correct value of $\rho$,
this will not be pursued here.  

\end{itemize}

\section{Uniform Approximation}

It is now easy to extend the results of the previous section to obtain results
for the uniform approximation of the oscillatory part of Riemann theta
functions. In Theorem 2, the sum extends over all terms whose summation index
vector is inside an ellipsoid. The size of this ellipsoid ($R$) is determined by
the allowed error of the pointwise approximation. The shape of the ellipsoid
depends on $\mbf{\Omega}$, but not on $\mbf{z}$. The center of the ellipsoid is
$-[[\Y^{-1}\y]]$. So, if different arguments $\mbf{z}$ are considered, only the
center of the ellipsoid changes and not its shape or size. But, this center can
only wander over a $g$-dimensional cube of volume 1, centered around the origin.
This leads to the following

\begin{theo}[Uniform Approximation]\la{theo:uniform}
The Riemann theta function is approximated by 
\beq
\theta(\mbf{z}|\mbf{\Omega})\approx e^{\pi \y\cdot \Y^{-1} \cdot \y}
\sum_{U_R} e^{2\pi
i\left(\frac{1}{2}\left(\n-\left[\Y^{-1}\y\right]\right)\cdot
\X\cdot\left(\n-\left[\Y^{-1}\y\right]\right)+
\left(\n-\left[\Y^{-1}\y\right]\right)\cdot\x\right)} e^{-||\mbf{v}(\n)||^2},
\eeq
\sloppypar \no 
with absolute error $\epsilon$ on the sum. The approximation of the sum 
is uniform in
$\mbf{z}$. Here $\v(\n)=\sqrt{\pi}\T(\n+[[\Y^{-1}\y]])$, 
\beq
U_R=\left\{\n\in \mathbb{Z}^g \big|\pi (\n-\mbf{c})\cdot \Y\cdot 
(\n -\mbf{c})<R^2, |c_j|<1/2, j=1, \ldots, g\right\}.
\eeq
Let $\Lambda$ $=$ 
$\left\{\sqrt{\pi}\T(\n+[[\Y^{-1}\y]])~|\n\in \mathbb{Z}^g\right\}$. 
The shortest
distance between any two points of $\Lambda$ is denoted by $\rho$. Then the 
radius
$R$ is determined as the greater of $(\sqrt{2g}+\rho)/2$ and the real positive
solution of $\epsilon=
g\,2^{g-1}\Gamma\left(g/2,(R-\rho/2)^2\right)/\rho^g$.  
\end{theo}

\begin{figure}[htb]
\centerline{\psfig{file=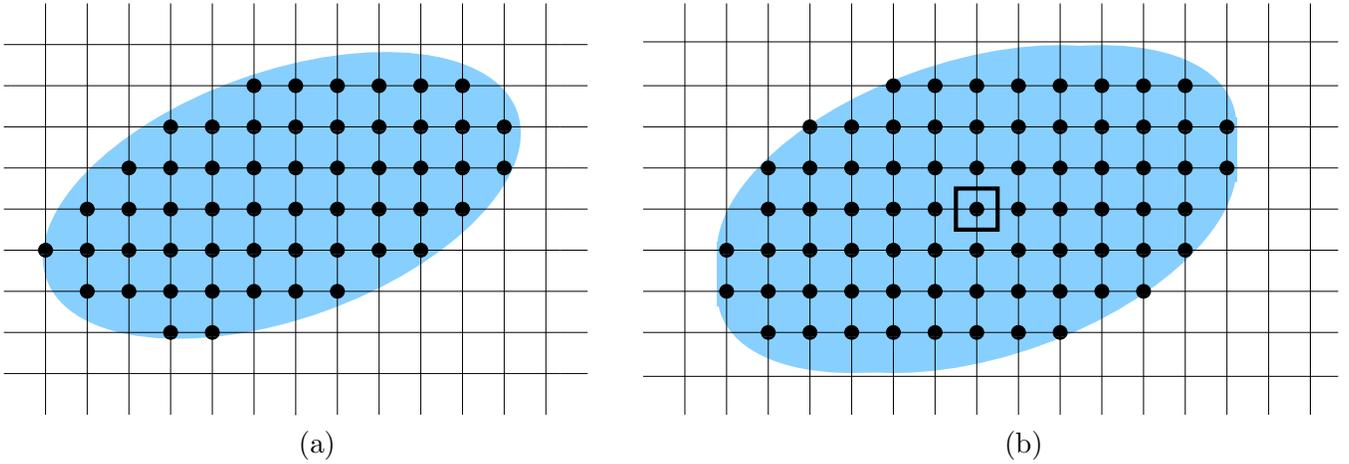,width=7in}} 
\centerline{(a)\hspace*{3.5in}(b)}
\caption{\label{fig:ellipsoids}
Summation indices (a) for a pointwise approximation, 
inside an ellipsoid centered at $[[\Y^{-1}\y]]$, and (b) for a uniform
approximation, inside an ellipsoid whose center moves around the unit cube.}
\end{figure}

Thus, to obtain a uniform in $\mbf{z}$ approximation with absolute error
$\epsilon$ for the oscillatory part of the Riemann theta function parametrized
by $\mbf{\Omega}$, it suffices to add all terms whose summation index is inside
the envelope of all ellipsoids of size $R$ (determined by $\epsilon$) whose
shape is determined by $\mbf{\Omega}$, and whose center is at any point inside
the unit cube centered at the origin. This object is also obtained as the
convex hull of all ellipsoids of size $R$ placed at all corners of the unit
cube centered at the origin. This situation is illustrated in Fig.
\ref{fig:ellipsoids}. The ellipsoid on the left illustrates the scenario of
obtaining the summation indices for a pointwise approximation. The deformed
ellipsoid on the right illustrates the case of a uniform approximation. The
deformation is a consequence of the wandering of the center of the ellipsoid
over the unit cube. 

It is clear that using a uniform approximation results in more terms of the sum
being used. In many applications (graphics, creation of value tables, $etc$.)
many values of the same Riemann theta function ($i.e.$, with constant
$\mbf{\Omega}$) are computed. In such cases it is beneficial to use the same
representation for the computation. 

\subsection*{Remarks}

\begin{itemize}

\item For the purposes of obtaining a uniform approximation formula, it is
essential that the exponential growth was factored out in \rf{rewrite}. 

\item The above theorem uses the 100 \% Error of Theorem \ref{theo:2}. Just as in the
case of a pointwise approximation, it is possible to consider a uniform
approximation using a Fill Factor Error. This results in similar modifications
to Theorem \ref{theo:uniform} as for Theorem \ref{theo:2}. 

\item Determining the set $U_R$ is not trivial. In practice, it may be more
convenient to work with a single ellipsoid centered at the origin, which is
too large. 

\end{itemize}

As an example, consider the Riemann matrix

\beq\la{rmatrix}
\mbf{\Omega}=\left(\ba{cc}
1.690983006+0.9510565162\,i& 1.5+0.363271264\,i\\
1.5+0.363271264\,i&1.309016994+0.9510565162\,i
\ea\right).
\eeq

\no This is the Riemann matrix associated with the genus 2 Riemann surface
obtained by compactifying and desingularizing the algebraic curve
$\mu^3-\lambda^7+2\lambda^3\mu=0$. It was computed using the algorithms
outlined in \cite{dvh1}. Using the above theorem with
$\epsilon=0.001$, 23 terms have a contribution to the oscillatory part of the
Riemann theta function. They have summation
indices
\bea\nonumber
U_R(100\%E)&=&
\left\{(-2,2),(2,-2), (0,2), (0,-2), (2,-1), (-2,1), 
(2,0), (-2,0), (0,-1),\right.\\\nonumber
&&(0,1),(-1,-1),(1,1),(1,0),(-1,0),(-2,-1),(2,1),(-1,1),
(1,-1),\\\la{ur1}
&&\left.(0,0),(-1,-2),(1,2),(-1,2),(1,-2)
\right\}.
\eea

\no Thus a sum of 23 terms suffices to approximate the oscillatory part of the
Riemann theta function associated with \rf{rmatrix} with an absolute error of
$\epsilon=0.001$. This approximation is valid for all $\mbf{z}=(x_1+i y_1, 
x_2+i y_2)$. These 23 terms should be compared to the 12-17 terms required for
a pointwise approximation. The pointwise approximation uses sets of summation
indices which are subsets of \rf{ur1}, but these subsets differ depending on
the value of  $\mbf{z}=(x_1+i y_1,  x_2+i y_2)$.

\begin{figure}[p]
\begin{tabular}{ccc}
\psfig{file=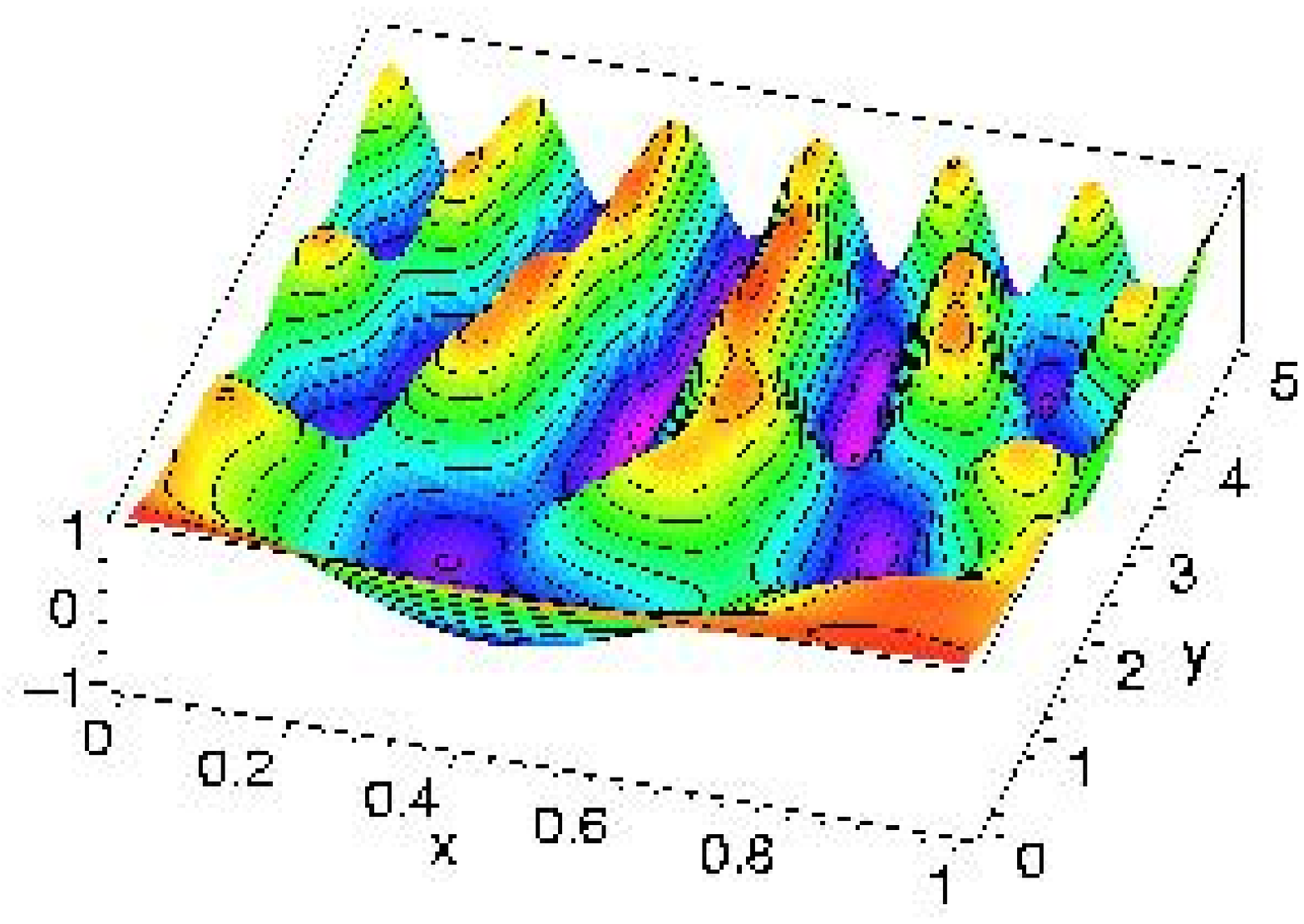,width=2in,angle=270}&
\psfig{file=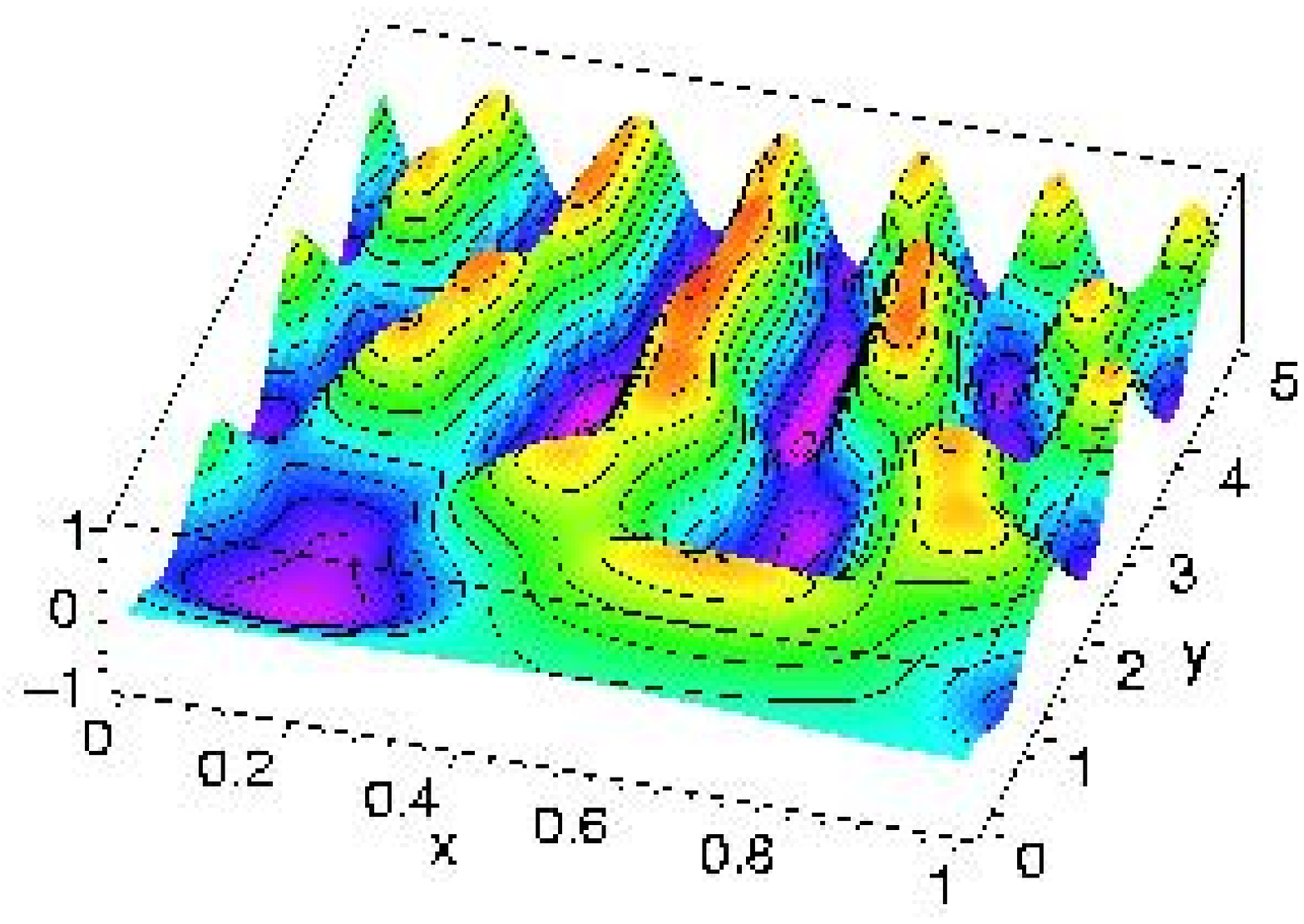,width=2in,angle=270}&
\psfig{file=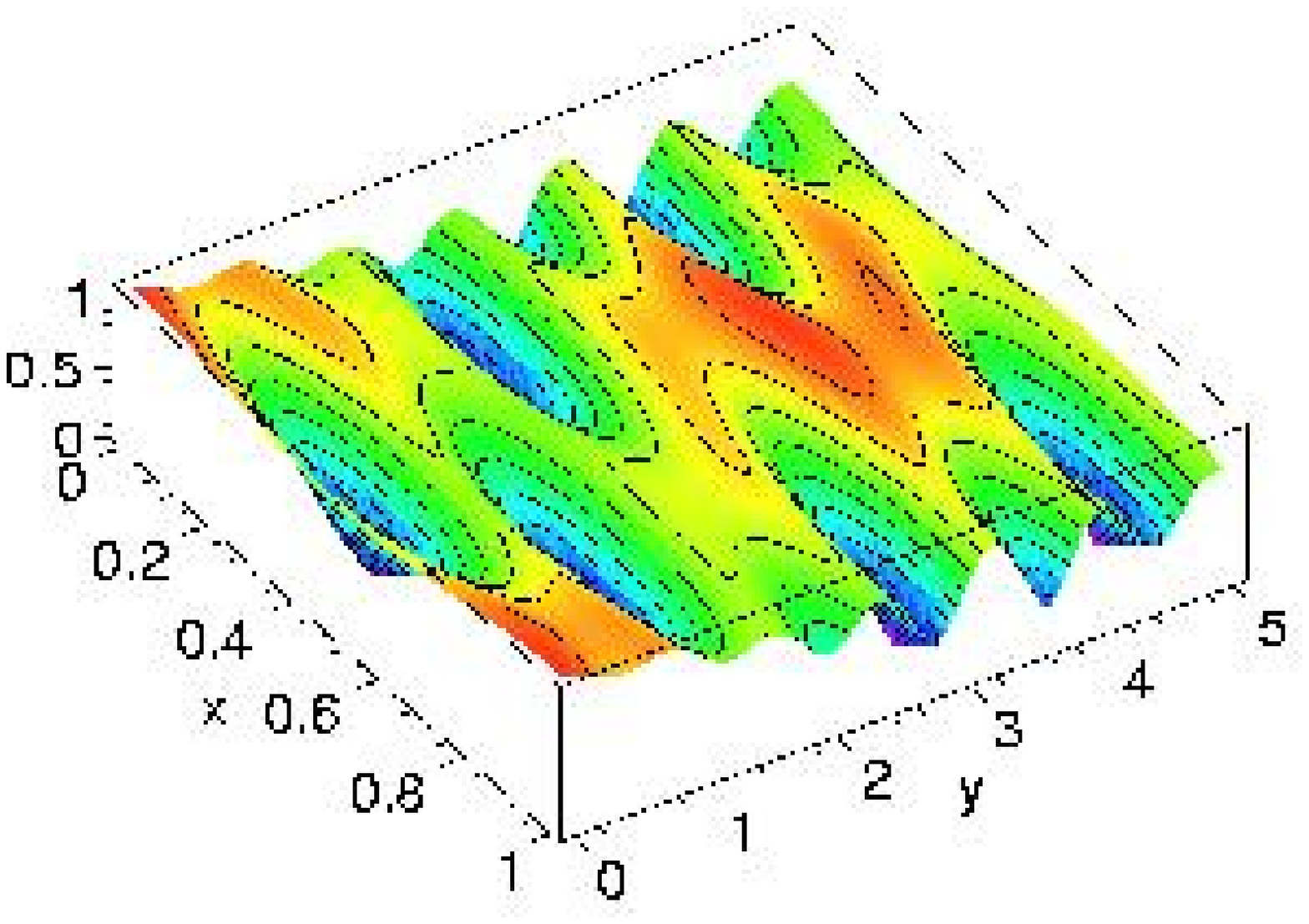,width=2in,angle=270}\\\vspace*{0.2in}
(a$_1$) & (b$_1$) & (c$_1$)\\
\psfig{file=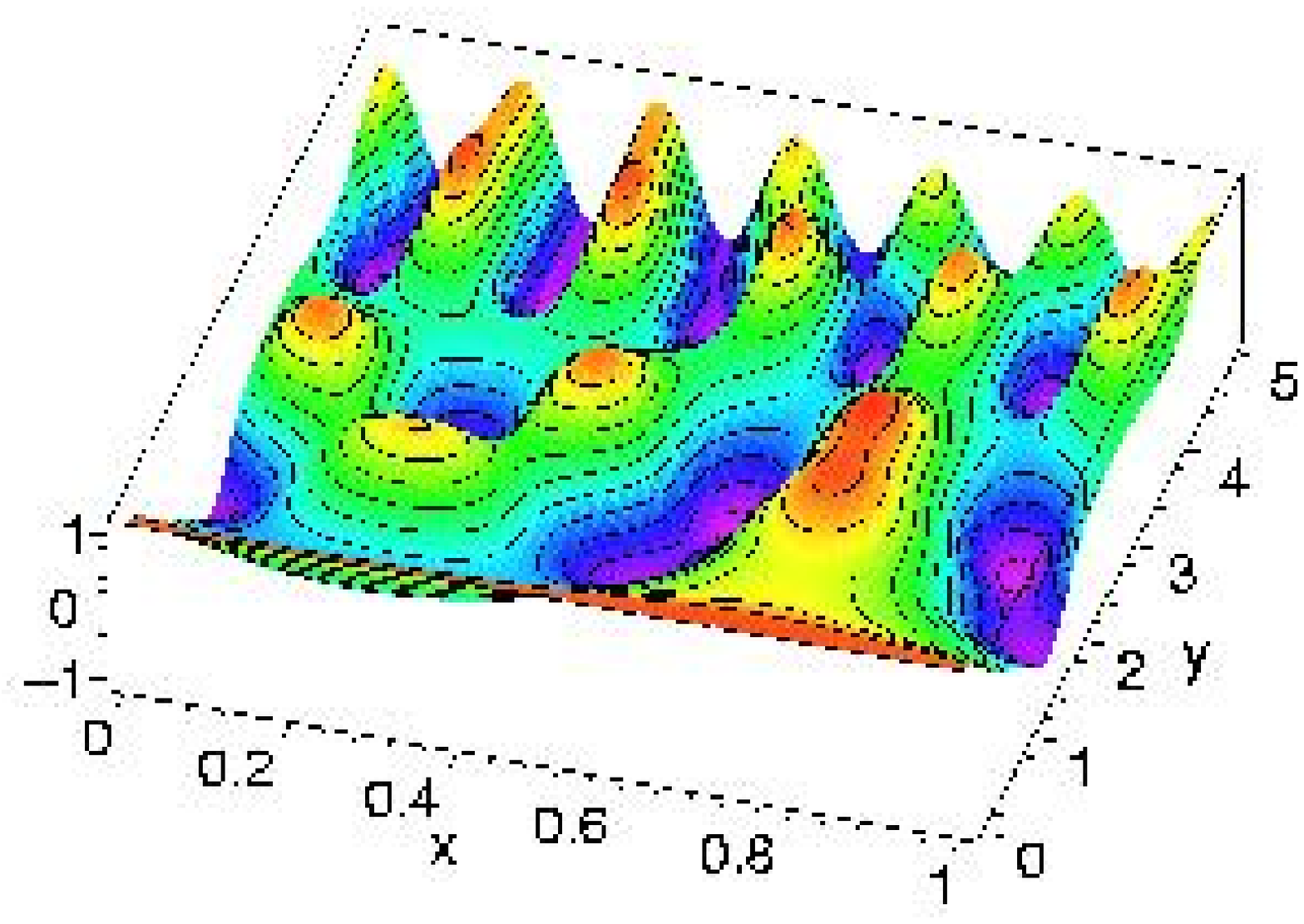,width=2in,angle=270}&
\psfig{file=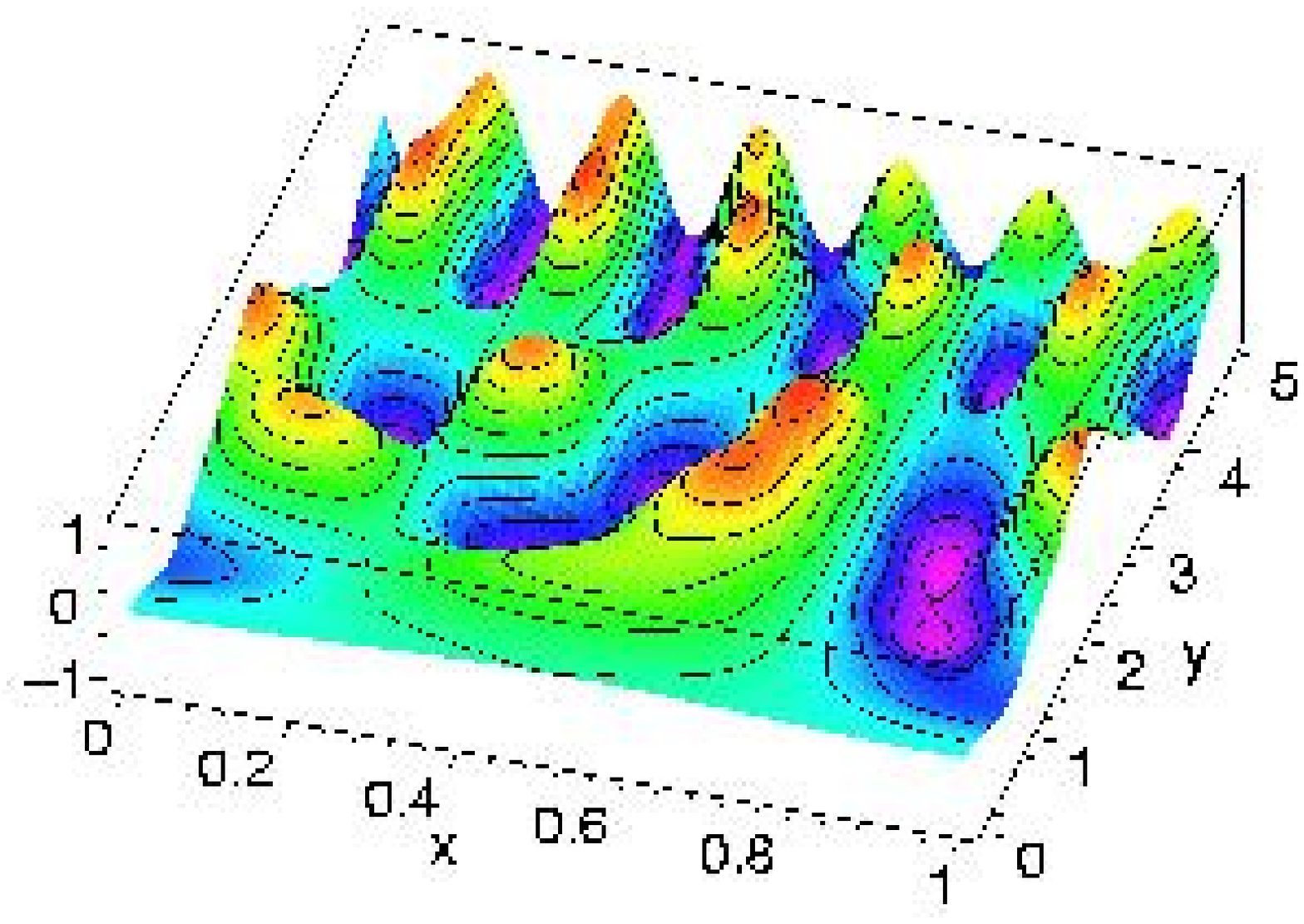,width=2in,angle=270}&
\psfig{file=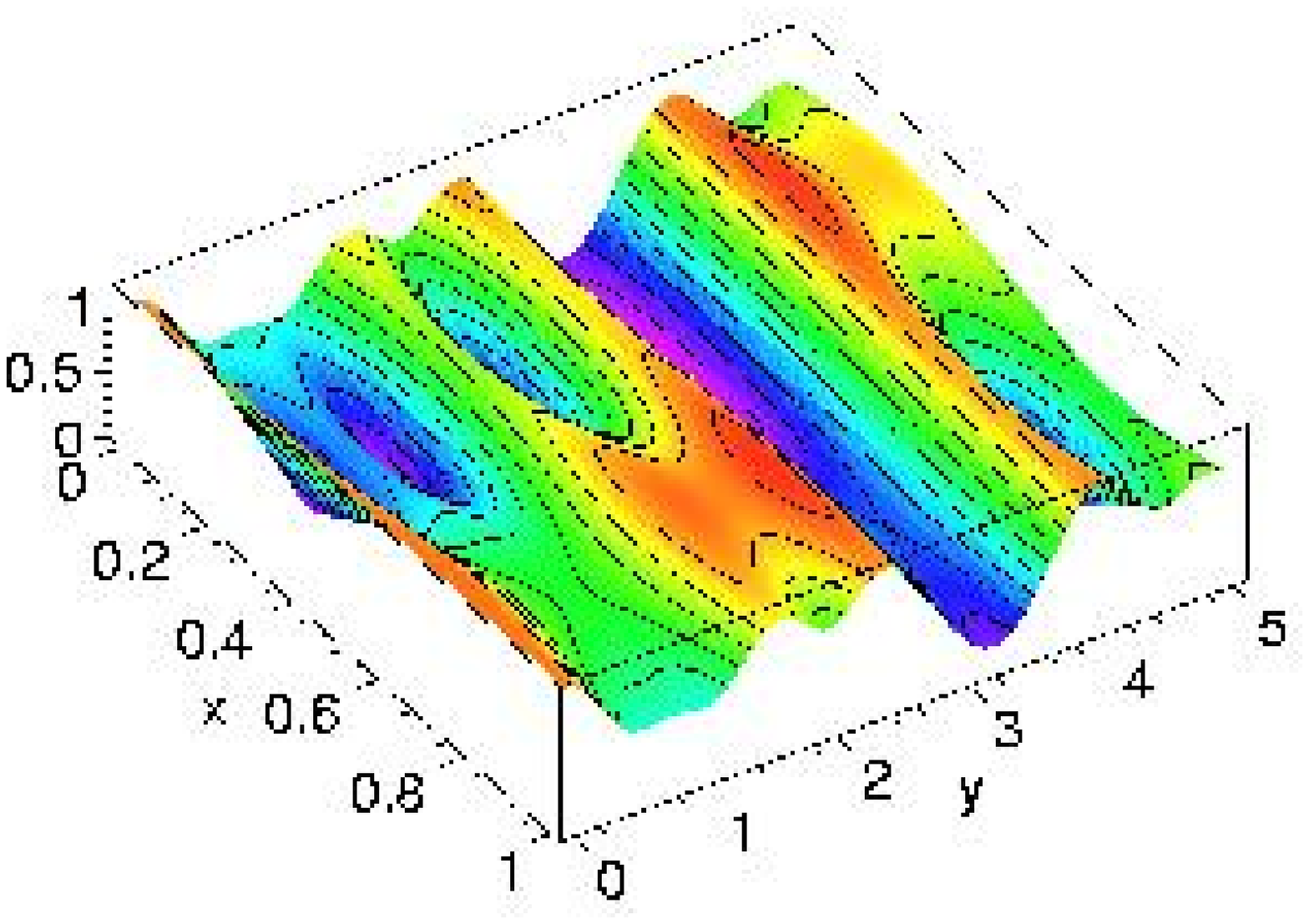,width=2in,angle=270}\\\vspace*{0.2in}
(a$_2$) & (b$_2$) & (c$_2$)\\
\psfig{file=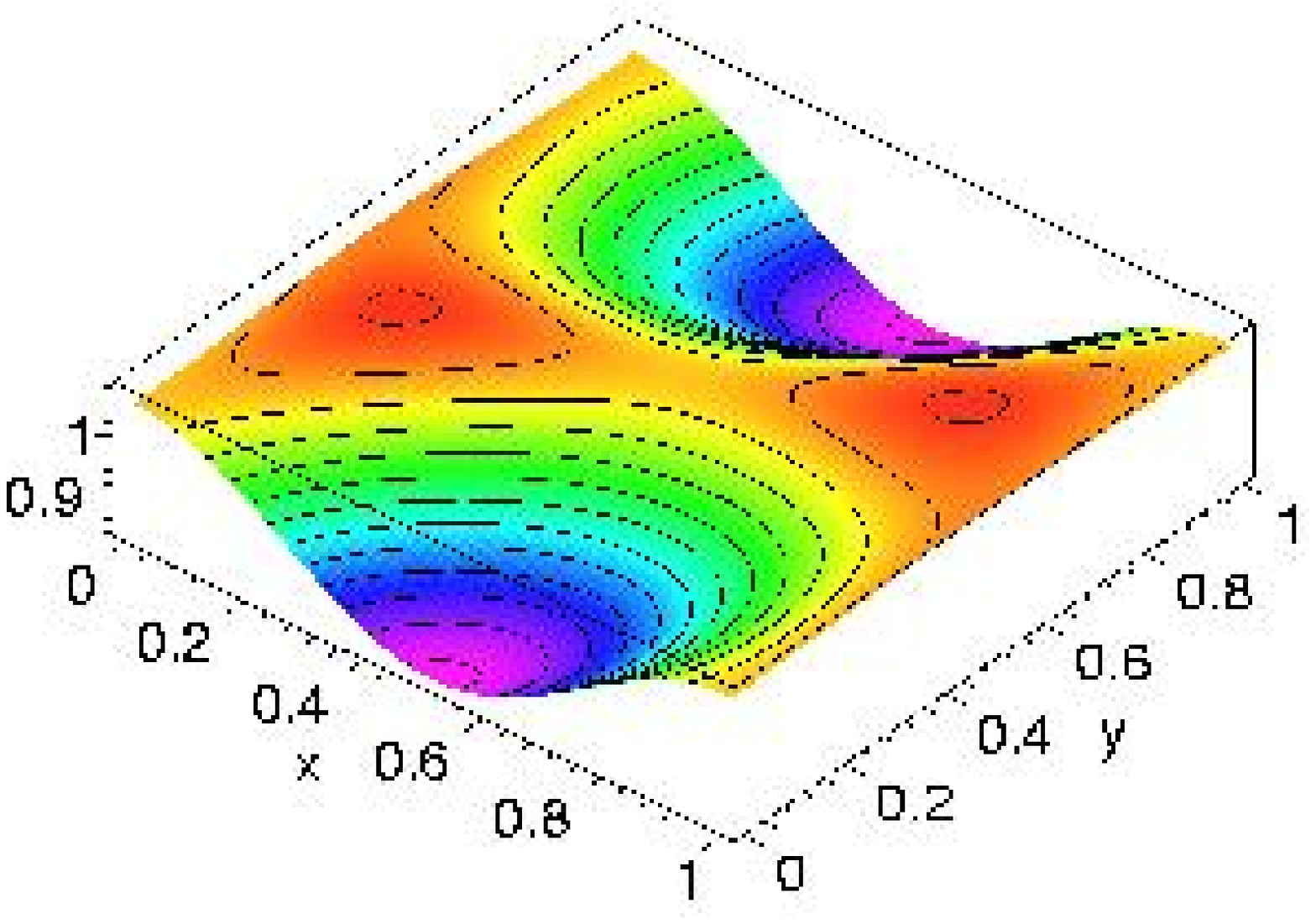,width=2in,angle=270}&
\psfig{file=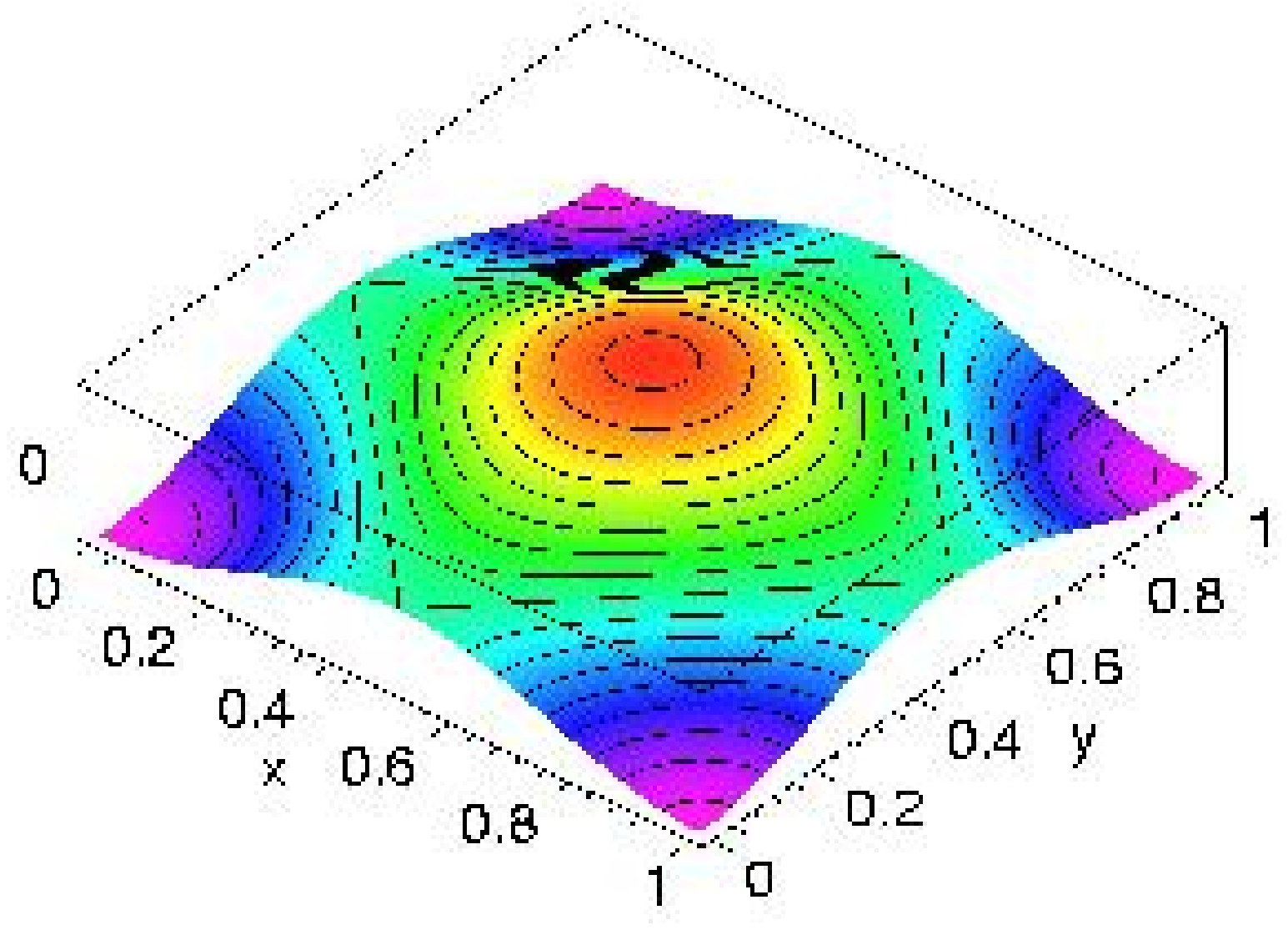,width=2in,angle=270}&
\psfig{file=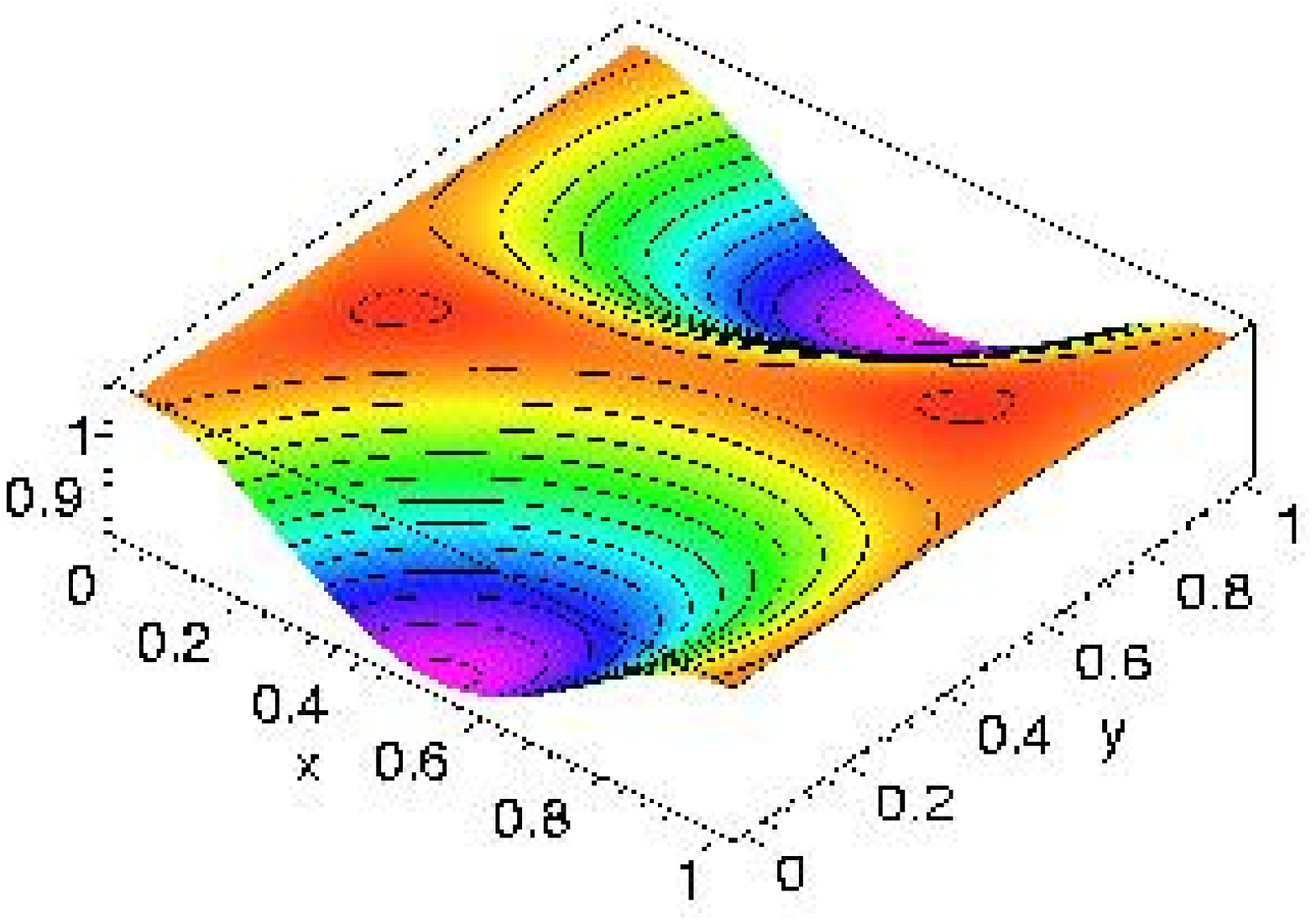,width=2in,angle=270}\\\vspace*{0.2in}
(a$_3$) & (b$_3$) & (c$_3$)\\
\psfig{file=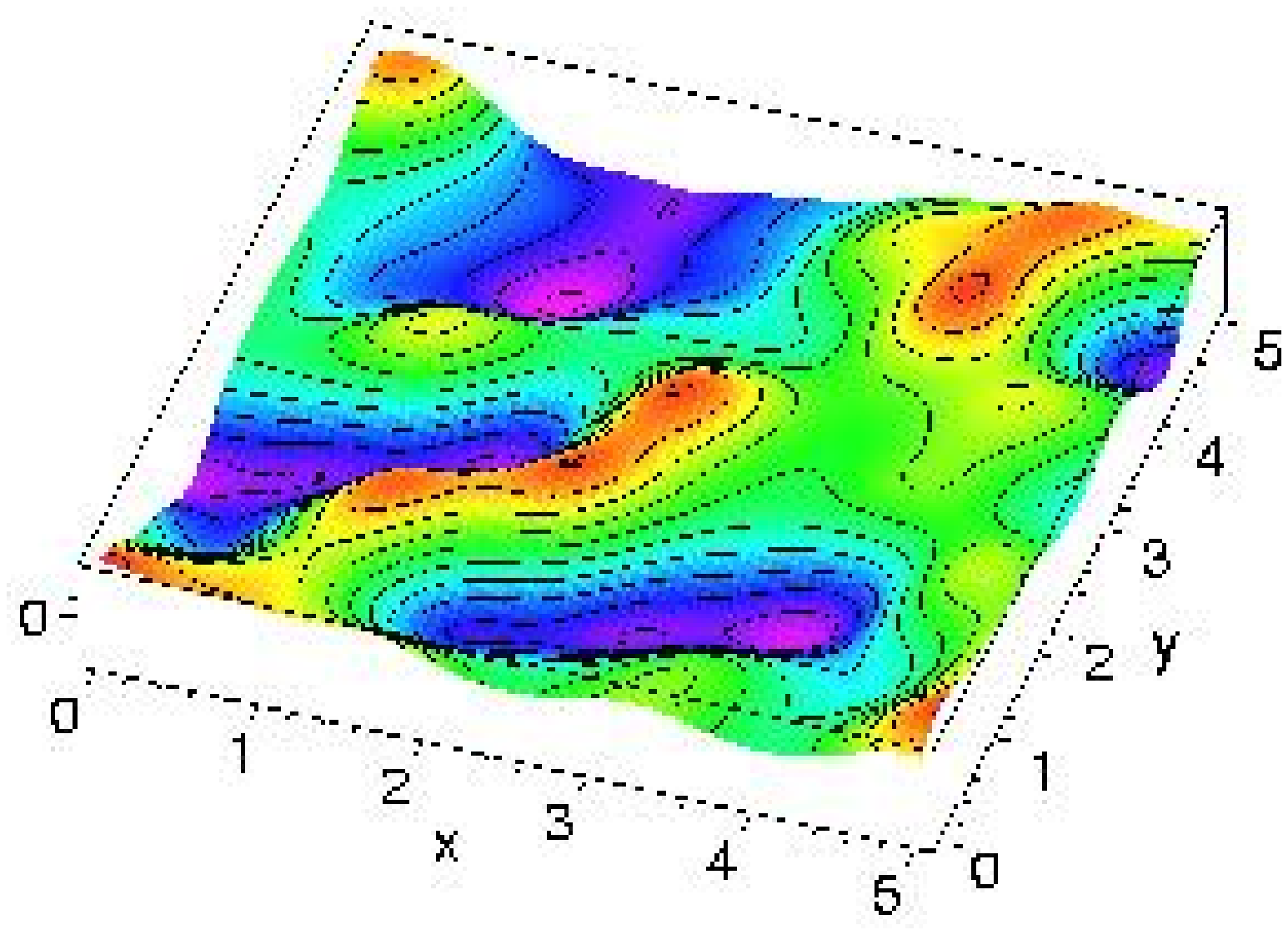,width=2in,angle=270}&
\psfig{file=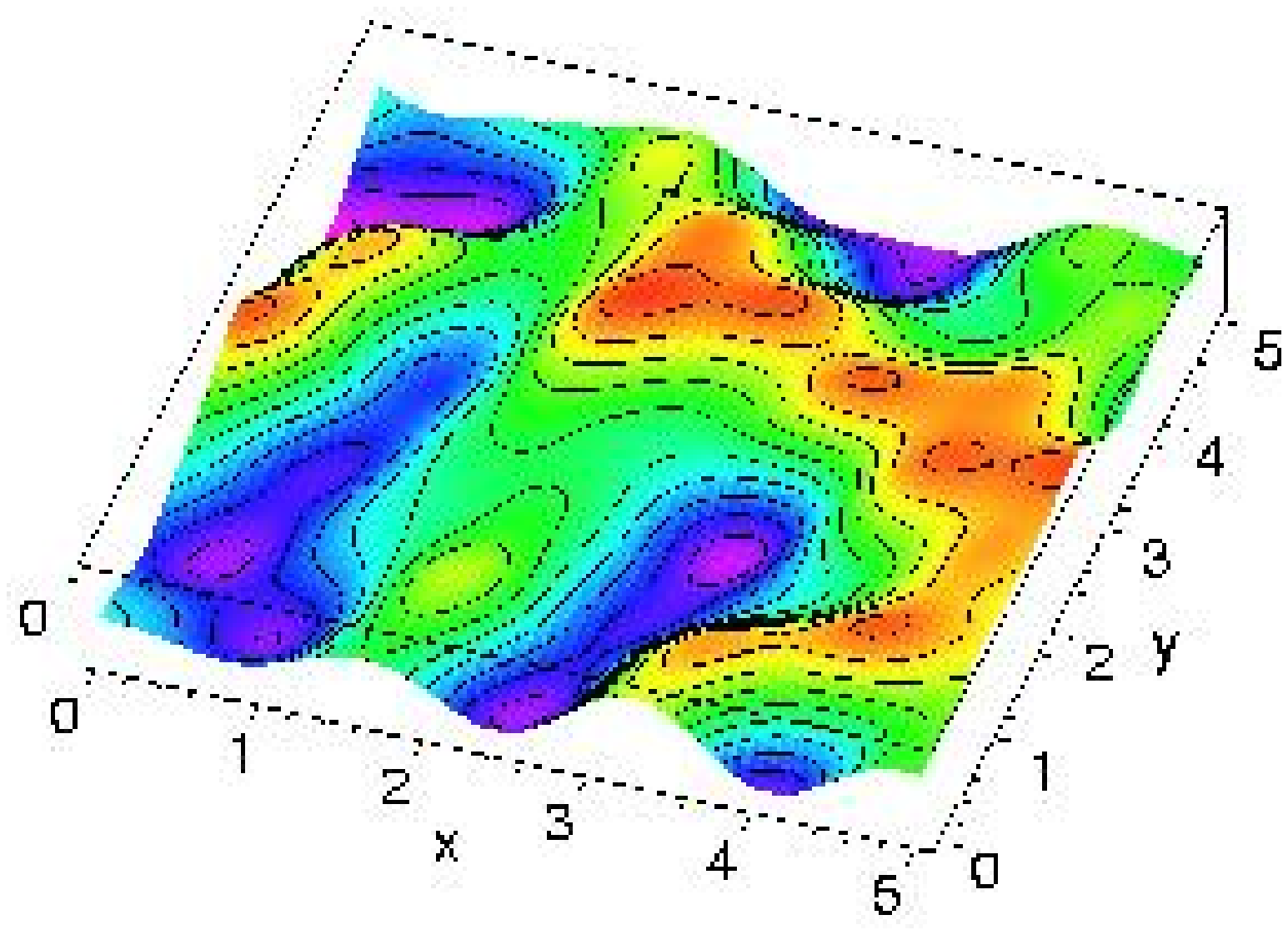,width=2in,angle=270}&
\psfig{file=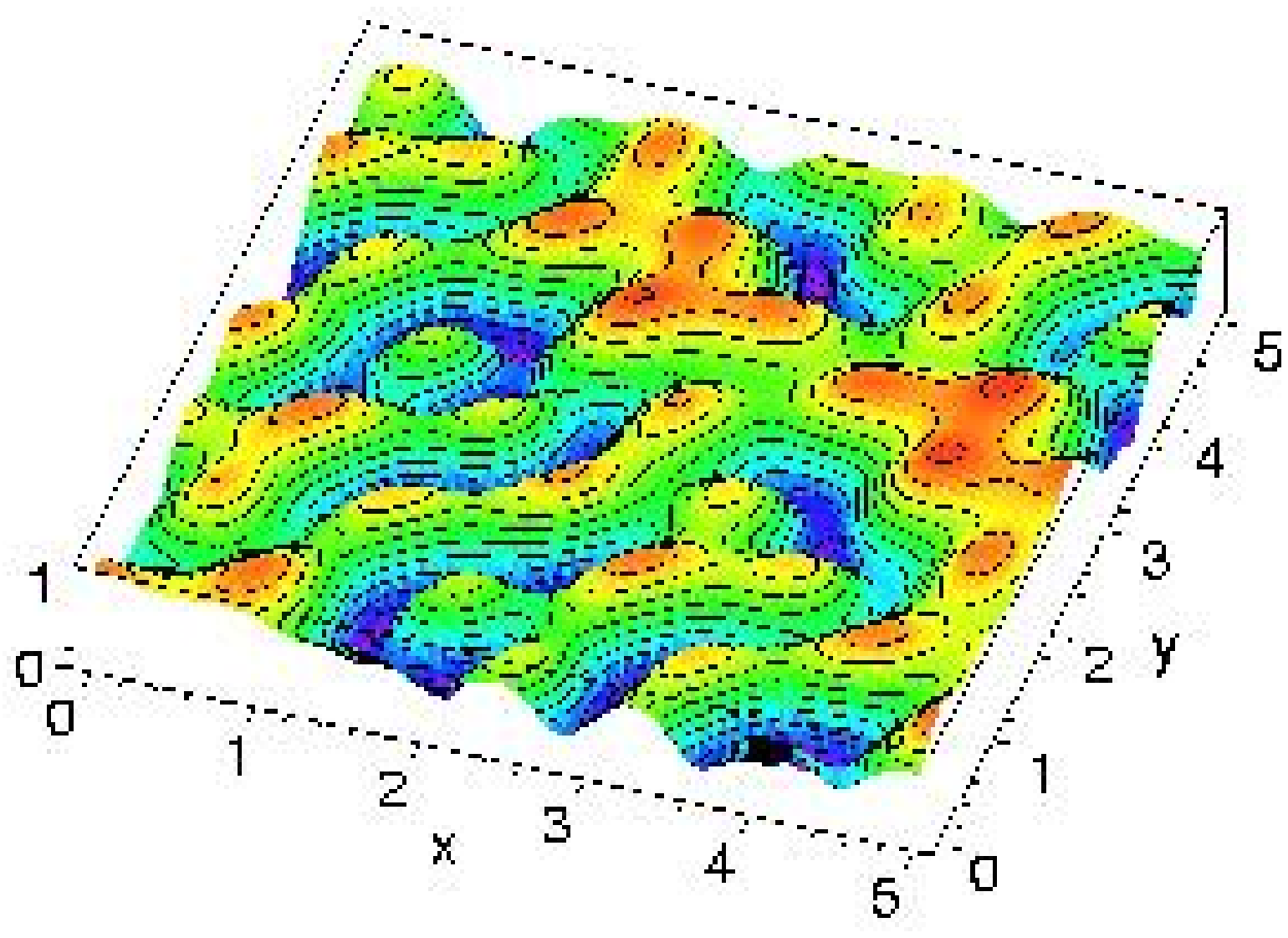,width=2in,angle=270}\\\vspace*{0.2in}
(a$_4$) & (b$_4$) & (c$_4$)
\end{tabular}
\caption{\label{fig:plots}
Various views of the oscillatory part of the Riemann theta function parametrized by the Riemann
matrix given in \rf{rmatrix}, with FFE=0.001. 
All plots are oscillatory parts of
$\theta(x+iy,0|\mbf{\Omega})$ (index 1), 
$\theta(0, x+iy|\mbf{\Omega})$ (index 2), 
$\theta(x,y|\mbf{\Omega})$ (index 3), 
$\theta(ix, iy|\mbf{\Omega})$ (index 4). Shown are the real part (a),
the imaginary part (b) and the absolute value (c).}
\end{figure}

The uniform approximation can be used to graph various
slices of the oscillatory part of the Riemann theta function. This is a complex
function of four real variables, of which many different slices are possible.
Twelve graphs are shown in Fig. \ref{fig:plots}, using the uniform
approximation. Note that the oscillatory part of the 
Riemann theta function is periodic with period 1 in the real part of all
components of $\mbf{z}$. This periodicity is inherited from the Riemann theta
function. 

\section{Derivatives of Riemann theta functions}

In this section, approximation results for derivatives of Riemann theta
functions are obtained. The results are not as clean as for the Riemann theta
function itself: even after removing the exponential growth, algebraic growth
terms remain in the sum. The order of the algebraic growth equals the order of
the derivative. There is only growth on the imaginary axes of the arguments 
$\mbf{z}$. Therefore, approximations of the sum with absolute error are bound
to be valid only pointwise in $\mbf{z}$. It is possible to obtain uniform
approximations in $\mbf{z}$ which are valid in a bounded area of $\mathbb{C}^g$.
We give the details for directional derivatives of first and second order.
Results for higher-order directional derivatives follow in a similar fashion. 

\subsection{First-order derivatives}

With $N=1$ and $\k^{(1)}=\k$, \rf{rewrited} gives
\bea\nonumber 
&&\!\!\!\!e^{-\pi \y\cdot\Y^{-1}\cdot\y}\,
D(\k)\theta(\mbf{z}|\mbf{\Omega})\\\nonumber
&=&
2\pi i\, \k \cdot \!\!
\sum_{\n\in \mathbb{Z}^g}\!\!\left(\n\!-\!\left[\Y^{-1}\y\right]
\right) e^{2\pi i\left(\frac{1}{2}\left(\n-\left[\Y^{-1}\y\right]\right)
\cdot \X\cdot\left(\n-\left[\Y^{-1}\y\right]\right)+
\left(\n-\left[\Y^{-1}\y\right]\right)\right)}e^{-||\v(\n)||^2}\\\nonumber
&=&\lim_{R\rightarrow\infty}
2\pi i\, \k \cdot \!\!\sum_{S_R}
\left(\n\!-\!\left[\Y^{-1}\y\right]
\right) e^{2\pi i\left(\frac{1}{2}\left(\n-\left[\Y^{-1}\y\right]\right)
\cdot \X\cdot\left(\n-\left[\Y^{-1}\y\right]\right)+
\left(\n-\left[\Y^{-1}\y\right]\right)\right)}\times\\
&&~~~~~~~~~~~~~~~~~~~~~~~~~~~~~~~~~~~~~~~~~~~~~~
~~~~~~~~~~~~~~~~~~~~~~~~~~~~~~~~~~~~~\times e^{-||\v(\n)||^2},
\eea
where $S_R$ and $\v(\n)$ are defined as in \rf{SR}. As before, let 
\bea\nonumber 
&&D_R(\k)\theta(\mbf{z}|\mbf{\Omega})
=e^{\pi \y\cdot\Y^{-1}\cdot \y} \times\\
&&\times
2\pi i\, \k \cdot \!\!\sum_{S_R}
\left(\n\!-\!\left[\Y^{-1}\y\right]
\right) e^{2\pi i\left(\frac{1}{2}\left(\n-\left[\Y^{-1}\y\right]\right)
\cdot \X\cdot\left(\n-\left[\Y^{-1}\y\right]\right)+
\left(\n-\left[\Y^{-1}\y\right]\right)\right)}e^{-||\v(\n)||^2},
\eea
then
\bea
\epsilon(R)&=&|D(\k)\theta(\mbf{z}|\mbf{\Omega})-
D_R(\k)\theta(\mbf{z}|\mbf{\Omega})|\,e^{\pi \y\cdot\Y^{-1}\cdot \y}\\
\nonumber
&=&\left|
2\pi i\, \k \cdot \!\!
\sum_{\Lambda\setminus S_R}\!\!\left(\n\!-\!\left[\Y^{-1}\y\right]
\right) e^{2\pi i\left(\frac{1}{2}\left(\n-\left[\Y^{-1}\y\right]\right)
\cdot \X\cdot\left(\n-\left[\Y^{-1}\y\right]\right)+
\left(\n-\left[\Y^{-1}\y\right]\right)\right)}e^{-||\v(\n)||^2}
\right|\\\nonumber
&\leq& 2\pi ||\k||\sum_{\Lambda\setminus S_R}
\left\|\n-\left[\Y^{-1}\y\right]\right\|\,e^{-||\v(\n)||^2}\\\nonumber
&=&2\pi ||\k||\sum_{\Lambda\setminus S_R}
\left\|\frac{1}{\sqrt{\pi}}\T^{-1}\v(\n)-\left[\left[\Y^{-1}\y\right]\right]-
\left[\Y^{-1}\y\right]\right\|\,e^{-||\v(\n)||^2}\\\nonumber
&=&2\pi ||\k||\sum_{\Lambda\setminus S_R}
\left\|\frac{1}{\sqrt{\pi}}\T^{-1}\v(\n)-\Y^{-1}\y\right\|\,e^{-||\v(\n)||^2}\\
\nonumber
&\leq&
2\pi ||\k||\sum_{\Lambda\setminus S_R}
\left\|\frac{1}{\sqrt{\pi}}\T^{-1}\v(\n)\right\|\,e^{-||\v(\n)||^2}+
2 \pi ||\k||\left\|\Y^{-1}\y\right\|\sum_{\Lambda\setminus S_R}
\,e^{-||\v(\n)||^2}\\\la{somenumber}
&\leq&
2\sqrt{\pi} ||\k|| \left\|\T^{-1}\right\|\sum_{\Lambda\setminus S_R}
\left\|\v(\n)\right\|\,e^{-||\v(\n)||^2}+
2 \pi ||\k||\left\|\Y^{-1}\y\right\|\sum_{\Lambda\setminus S_R}
\,e^{-||\v(\n)||^2},
\eea
where the Cauchy-Schwarz and triangle inequalities were used. The second term in
\rf{somenumber} is estimated as before. For the sum in the first term, Lemma
\ref{lemma:1} is used with $p=1$. However, this only applied in the region where
$||\v(\n)||e^{-||\v(\n)||^2}$ is subharmonic. According to Theorem \ref{theo:1},
this holds when $R>\frac{1}{2}\sqrt{g+2+\sqrt{g^2+8}}$. Thus
\beq\la{epsd}
\epsilon(R)\leq
\sqrt{\pi} g \|\k\|\left(\frac{2}{\rho}\right)^g
\left(\left\|\T^{-1}\right\|
\Gamma\left(\frac{g+1}{2},(R-\rho/2)^2\right)+
\sqrt{\pi}
\left\|\Y^{-1}\y\right\|
\Gamma\left(\frac{g}{2},(R-\rho/2)^2\right)\right),
\eeq
provided $R>\frac{1}{2}\left(\sqrt{g+2+\sqrt{g^2+8}}+\rho\right)$. 

\begin{theo}[Pointwise Approximation of the First Directional Derivative]
\la{theta:pointder} 
\sloppypar The directional derivative of the Riemann theta function
along $\k$ is approximated by 

\bea\nonumber
\!\!\!\!D(\k)\!\!\!\!\!\!\!\!\!\!&&
\theta(\mbf{z}|\Omega)=e^{\pi \y\cdot\Y^{-1}\cdot\y}\times\\
&&\times
2\pi i \k \cdot \!\!\sum_{S_R} \left(\n\!-\!\left[\Y^{-1}\y\right]
\right) e^{2\pi i\left(\frac{1}{2}\left(\n-\left[\Y^{-1}\y\right]\right)
\cdot \X\cdot\left(\n-\left[\Y^{-1}\y\right]\right)+
\left(\n-\left[\Y^{-1}\y\right]\right)\right)}e^{-||\v(\n)||^2},
\eea

\no with absolute error $\epsilon$ on the scalar product of $2\pi i \,\k$ with the sum. 
Here $S_R=\left\{\v(\n)\in
\Lambda\big|~||\v(n)||<R\right\}$,
$\Lambda$ $=$ 
$\left\{\sqrt{\pi}\T(\n+[[\Y^{-1}\y]])~|\n\in \mathbb{Z}^g\right\}$. The shortest
distance between any two points of $\Lambda$ is denoted by $\rho$. Then the 
radius
$R$ is determined as the greater of 
$\left(\sqrt{g+2+\sqrt{g^2+8}}+\rho\right)/2$ 
and the real positive
solution of $\epsilon=\sqrt{\pi} g \|\k\|\left(\frac{2}{\rho}\right)^g
\left(\left\|\T^{-1}\right\|
\Gamma\left(\frac{g+1}{2},(R-\rho/2)^2\right)+
\sqrt{\pi}
\left\|\Y^{-1}\y\right\|
\Gamma\left(\frac{g}{2},(R-\rho/2)^2\right)\right)
$. 
\end{theo}

This theorem clearly gives a pointwise approximation: not only is the location
of the ellipsoid $S_R$ dependent on the evaluation point $\mbf{z}$; more
importantly its size $R$ depends on it. This is the reason why this result
cannot be extended to a uniform in $\mbf{z}$ approximation theorem, as in the
case of Theorem \ref{theo:uniform}. By limiting the domain of $\y$, it is
possible to obtain a uniform in $\mbf{z}$ approximation theorem, valid over this
domain: 

\begin{theo}[Uniform Approximation of the First Directional
Derivative]\la{theo:unifder}
\sloppypar The directional derivative of the Riemann theta function
along $\k$ is approximated by 

\bea\nonumber
\!\!\!\!D(\k)\!\!\!\!\!\!\!\!\!\!&&
\theta(\mbf{z}|\Omega)=e^{\pi \y\cdot\Y^{-1}\cdot\y}\times\\
&&\times
2\pi i \k \cdot \!\!\sum_{U_R} \left(\n\!-\!\left[\Y^{-1}\y\right]
\right) e^{2\pi i\left(\frac{1}{2}\left(\n-\left[\Y^{-1}\y\right]\right)
\cdot \X\cdot\left(\n-\left[\Y^{-1}\y\right]\right)+
\left(\n-\left[\Y^{-1}\y\right]\right)\right)}e^{-||\v(\n)||^2},
\eea

\no with absolute error $\epsilon$ on the scalar product of $2\pi i \,\k$ with the sum,
where the approximation is uniform in $\mbf{z}$, $\mbf{z}=\x+i\y$, for $\y\in
B_L^g(0)$. Here $\v(\n)=\sqrt{\pi}\T(\n+[[\Y^{-1}\y]])$, and 
\beq
U_R=\left\{\n\in \mathbb{Z}^g \big|\pi (\n-\mbf{c})\cdot \Y\cdot 
(\n -\mbf{c})<R^2, |c_j|<1/2, j=1, \ldots, g\right\}.
\eeq
Let $\Lambda$ $=$ 
$\left\{\sqrt{\pi}\T(\n+[[\Y^{-1}\y]])~|\n\in \mathbb{Z}^g\right\}$. The shortest
distance between any two points of $\Lambda$ is denoted by $\rho$. Then the 
radius
$R$ is determined as the greater of 
$\left(\sqrt{g+2+\sqrt{g^2+8}}+\rho\right)/2$ 
and the real positive
solution of $\epsilon$ $=$ $\sqrt{\pi} g \|\k\| \left\|\T^{-1}\right\|
\left(\frac{2}{\rho}\right)^g$
$\left(
\Gamma\left(\frac{g+1}{2},(R-\rho/2)^2\right)+
\sqrt{\pi} L 
\left\|\T^{-1}\right\|
\Gamma\left(\frac{g}{2},(R-\rho/2)^2\right)\right)
$. 
\end{theo}

\no{\bf Proof.} This follows easily from the Pointwise Approximation result, by
using $\|\Y^{-1}\y\|\leq \|\Y^{-1}\|\|\y\|$ and $\|\Y^{-1}\|\leq \|\T^{-1}\|^2$.
\hfill $\bbox$

\vspace*{0.3in}

\begin{figure}[htb]
\begin{tabular}{cc}
\psfig{file=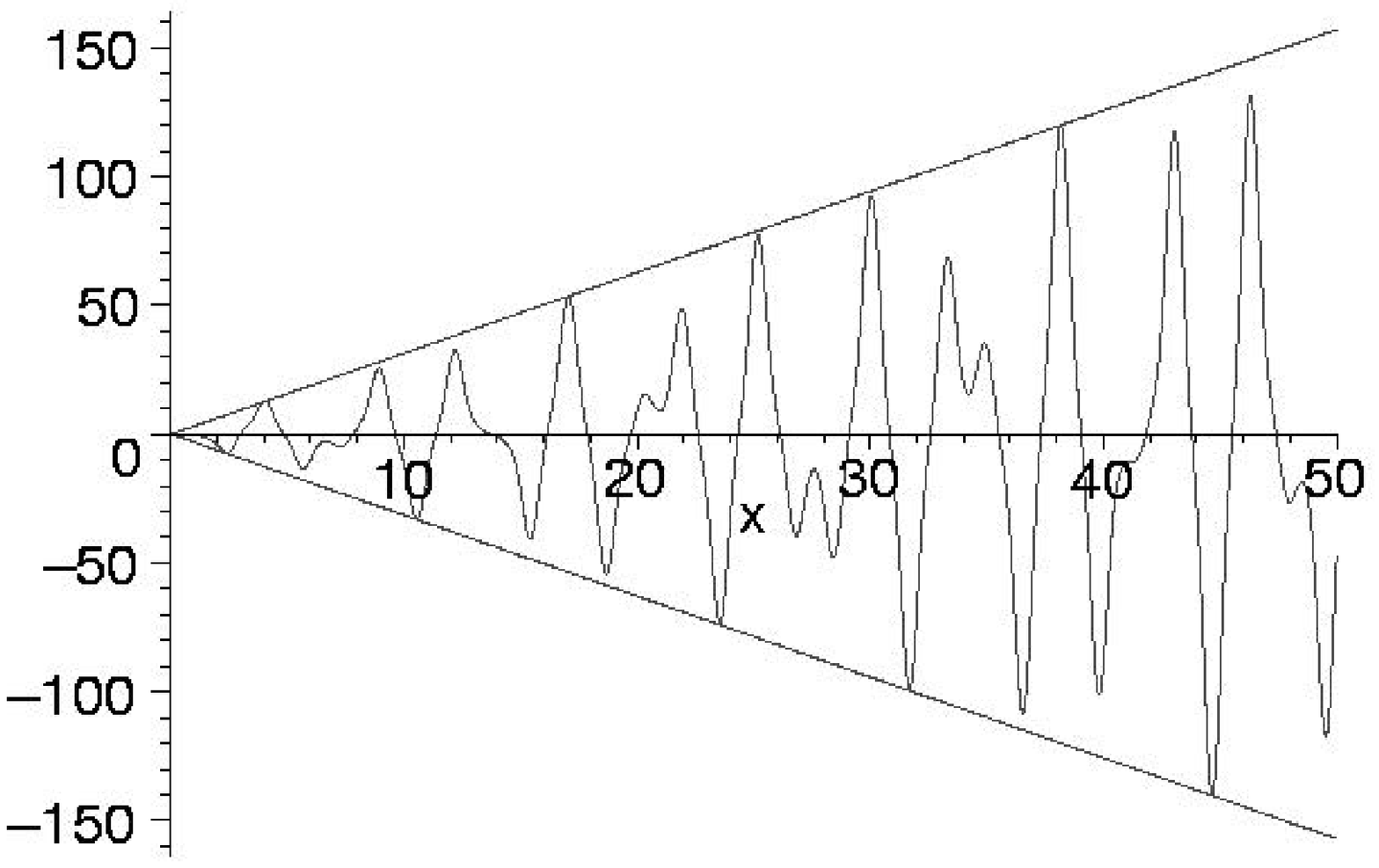,width=3in,angle=270}&
\psfig{file=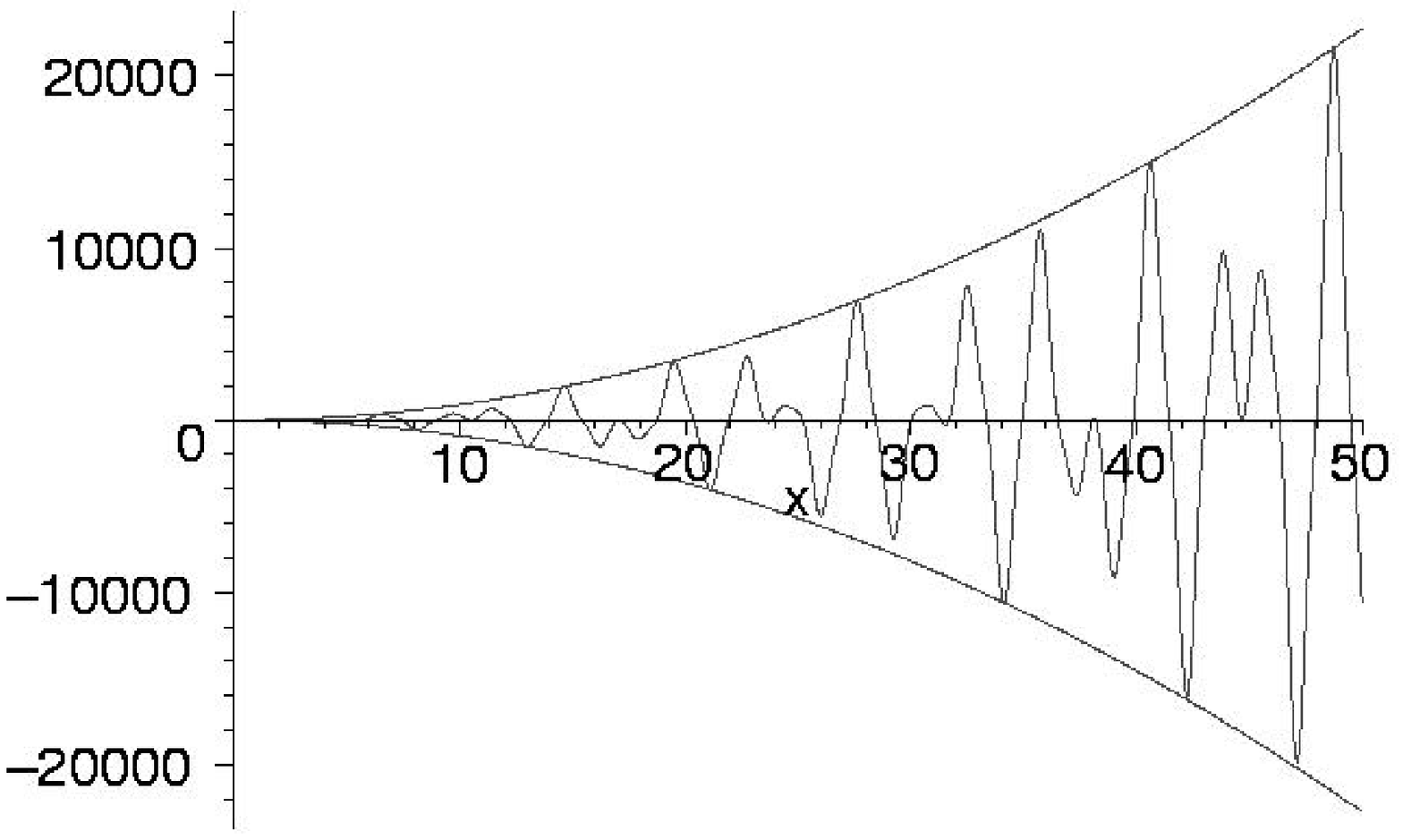,width=3in,angle=270}\\
(a) & (b)
\end{tabular}
\caption{\label{fig:growth}
Two directional derivatives of the Riemann theta function
$\theta(0,ix|\Omega)$ with $\Omega$ as in \rf{rmatrix}. As before, the
exponential growth-factor has been factored out. In (a) the directional
derivative of order one with $\k=(1,0)$ is shown. Fig. (b) illustrates the
quadratic growth of the directional derivative of order two, with
$\k^{(1)}=\k^{(2)}=(1,0)$. }
\end{figure}

This theorem is not as useful as Theorem \ref{theo:uniform}, due to the
restriction on the size of $\|\y\|$. When it is used, it results in a situation
where many terms are included which are only relevant for the evaluation of
values of the directional derivative near $\|\y\|=L$. However, for graphing or
other purposes where many evaluations of $D(\k)\theta(\mbf{z}|\mbf{\Omega})$
are required, centered around an area for $\mbox{z}$ where $\y=0$, such a
uniform approximation is usually beneficial compared to the pointwise
approximation. It was used in Fig. \ref{fig:growth}a to illustrate the
linear growth of the derivative of the Riemann theta function parametrized by
\rf{rmatrix}, after removal of the
exponential growth. 

\subsection{Second-order derivatives}

With $N=2$, \rf{rewrited} gives
\bea\nonumber 
&&\!\!\!\!e^{-\pi \y\cdot\Y^{-1}\cdot\y}\,
D(\k^{(1)},\k^{(2)})\theta(\mbf{z}|\mbf{\Omega})\\\nonumber
&=&
(2\pi i)^2
\sum_{\n\in \mathbb{Z}^g}
\left(\k^{(1)}\cdot\left(\n\!-\!\left[\Y^{-1}\y\right]\right)\right)
\left(\k^{(2)}\cdot\left(\n\!-\!\left[\Y^{-1}\y\right]\right)\right)\times\\
\nonumber
&&~~~~~~~~~~
\times e^{2\pi i\left(\frac{1}{2}\left(\n-\left[\Y^{-1}\y\right]\right)
\cdot \X\cdot\left(\n-\left[\Y^{-1}\y\right]\right)+
\left(\n-\left[\Y^{-1}\y\right]\right)\right)}e^{-||\v(\n)||^2}\\\nonumber
&=&(2\pi i)^2 \lim_{R\rightarrow\infty}
\sum_{S_R}
\left(\k^{(1)}\cdot\left(\n\!-\!\left[\Y^{-1}\y\right]\right)\right)
\left(\k^{(2)}\cdot\left(\n\!-\!\left[\Y^{-1}\y\right]\right)\right)\times
\\
&&~~~~~~~~~~~~~~~~~
\times e^{2\pi i\left(\frac{1}{2}\left(\n-\left[\Y^{-1}\y\right]\right)
\cdot \X\cdot\left(\n-\left[\Y^{-1}\y\right]\right)+
\left(\n-\left[\Y^{-1}\y\right]\right)\right)}e^{-||\v(\n)||^2}
\eea
where $S_R$ and $\v(\n)$ are defined as in \rf{SR}. Using similar steps as
before, and using Lemmas \ref{lemma:1} and \ref{lemma:2} with $p=2$ results in
the following theorems: 

\begin{theo}[Pointwise Approximation of the Second Directional Derivative]
\la{theta:pointderder} 
\sloppypar The directional derivative of second order 
along $\k^{(1)}$ and $\k^{(2)}$ 
of the Riemann theta function
is approximated by 

\bea\nonumber
D(\k^{(1)},\k^{(2)})\!\!\!\!\!\!\!\!\!\!&&
\theta(\mbf{z}|\mbf{\Omega})=e^{\pi \y\cdot\Y^{-1}\cdot\y}\times\\\nonumber
&&\times (2\pi i)^2 
\sum_{S_R}
\left(\k^{(1)}\cdot\left(\n\!-\!\left[\Y^{-1}\y\right]\right)\right)
\left(\k^{(2)}\cdot\left(\n\!-\!\left[\Y^{-1}\y\right]\right)\right)\times
\\
&&
\times e^{2\pi i\left(\frac{1}{2}\left(\n-\left[\Y^{-1}\y\right]\right)
\cdot \X\cdot\left(\n-\left[\Y^{-1}\y\right]\right)+
\left(\n-\left[\Y^{-1}\y\right]\right)\right)}e^{-||\v(\n)||^2}
\eea

\no with absolute error $\epsilon$ on the product of $(2\pi i)^2$ with the
sum.  Here $S_R=\left\{\v(\n)\in \Lambda\big|~||\v(n)||<R\right\}$, $\Lambda$
$=$  $\left\{\sqrt{\pi}\T(\n+[[\Y^{-1}\y]])~|\n\in \mathbb{Z}^g\right\}$. The
shortest distance between any two points of $\Lambda$ is denoted by $\rho$.
Then the  radius
$R$ is determined as the greater of 
$\left(\sqrt{g+4+\sqrt{g^2+16}}+\rho\right)/2$ 
and the real positive
solution of $\epsilon$ $=$ 
$2\pi g \|\k^{(1)}\|\,\|\k^{(2)}\|$
$\left(\frac{2}{\rho}\right)^g$
$\left(
\left\|\T^{-1}\right\|^2
\Gamma\left(\frac{g+2}{2},(R-\rho/2)^2\right)\right.$ $+$
$2\sqrt{\pi}\left\|\T^{-1}\right\|
\left\|\Y^{-1}\y\right\|
\Gamma\left(\frac{g+1}{2},(R-\rho/2)^2\right)$ $+$
$\left.\pi\left\|\Y^{-1}\y\right\|^2\Gamma\left(\frac{g}{2},(R-\rho/2)^2\right)
\right)
$. 
\end{theo}

\begin{theo}[Uniform Approximation of the Second Directional
Derivative]\la{theo:unifderder} 
The directional derivative of second order 
along $\k^{(1)}$ and $\k^{(2)}$ 
of the Riemann theta function
is approximated by 
\bea\nonumber
D(\k^{(1)},\k^{(2)})\!\!\!\!\!\!\!\!\!\!&&
\theta(\mbf{z}|\mbf{\Omega})=e^{\pi \y\cdot\Y^{-1}\cdot\y}\times\\\nonumber
&&\times (2\pi i)^2 
\sum_{U_R}
\left(\k^{(1)}\cdot\left(\n\!-\!\left[\Y^{-1}\y\right]\right)\right)
\left(\k^{(2)}\cdot\left(\n\!-\!\left[\Y^{-1}\y\right]\right)\right)\times
\\
&&
\times e^{2\pi i\left(\frac{1}{2}\left(\n-\left[\Y^{-1}\y\right]\right)
\cdot \X\cdot\left(\n-\left[\Y^{-1}\y\right]\right)+
\left(\n-\left[\Y^{-1}\y\right]\right)\right)}e^{-||\v(\n)||^2}
\eea
\no with absolute error $\epsilon$ on the scalar product of $(2\pi i)^2$ 
with the sum,
where the approximation is uniform in $\mbf{z}$, $\mbf{z}=\x+i\y$, for $\y\in
B_L^g(0)$. Here $\v(\n)=\sqrt{\pi}\T(\n+[[\Y^{-1}\y]])$, and 
\beq
U_R=\left\{\n\in \mathbb{Z}^g \big|\pi (\n-\mbf{c})\cdot \Y\cdot 
(\n -\mbf{c})<R^2, |c_j|<1/2, j=1, \ldots, g\right\}.
\eeq
Let $\Lambda$ $=$ 
$\left\{\sqrt{\pi}\T(\n+[[\Y^{-1}\y]])~|\n\in \mathbb{Z}^g\right\}$. The shortest
distance between any two points of $\Lambda$ is denoted by $\rho$. Then the 
radius
$R$ is determined as the greater of 
$\left(\sqrt{g+4+\sqrt{g^2+16}}+\rho\right)/2$ 
and the real positive
solution of $\epsilon$ $=$ 
$2\pi g \|\k^{(1)}\|\,\|\k^{(2)}\|$
$\left(\frac{2}{\rho}\right)^g$ $\left\|\T^{-1}\right\|^2$
$\left(
\Gamma\left(\frac{g+2}{2},(R-\rho/2)^2\right)\right.$ $+$
$2\sqrt{\pi}$ $\left\|\T^{-1}\right\| L
\Gamma\left(\frac{g+1}{2},(R-\rho/2)^2\right)$ $+$
$\left.\pi L^2 \left\|\T^{-1}\right\|^2\Gamma\left(\frac{g}{2},(R-\rho/2)^2\right)
\right)
$. 
\end{theo}

The same remarks as for the first-order directional derivative are valid here
as well. The last theorem has limited use due to its  restriction on the size
of $\|\y\|$.  It was used in Fig. \ref{fig:growth}b to illustrate the quadratic
growth of the derivative of the Riemann theta function parametrized by
\rf{rmatrix}, after removal of the exponential growth. 

\section{Siegel transformations}\la{sec:siegel}

Consider the Riemann theta function parametrized by the Riemann matrix

\beq\la{rm2}
\mbf{\Omega}=\frac{-1}{2\pi i}\left(\ba{cc}111.207 & 96.616\\
96.616 & 83.943\ea\right).
\eeq

\begin{figure}[htb]
\begin{tabular}{cc}
\psfig{file=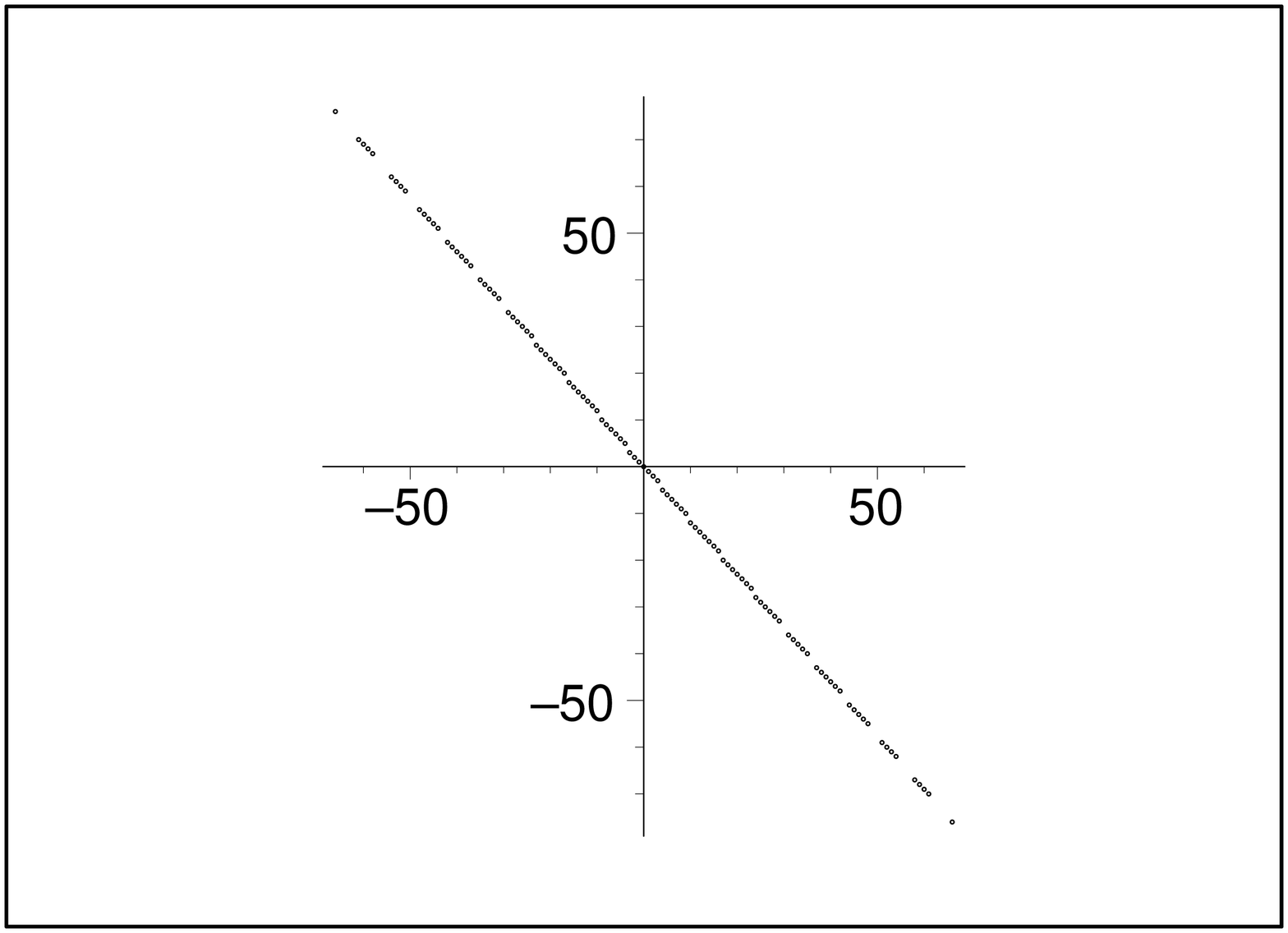,width=3in,angle=270}&
\psfig{file=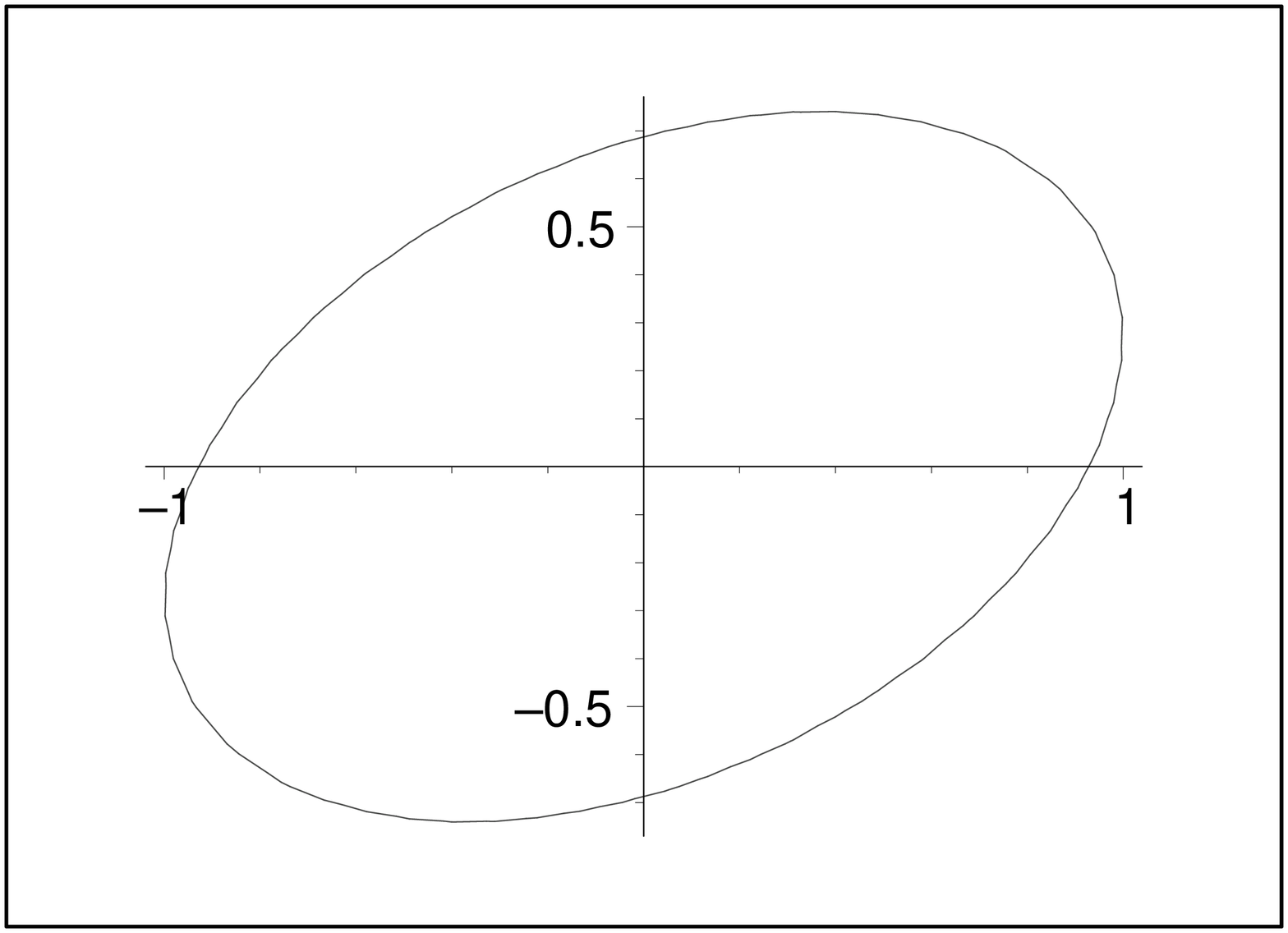,width=3in,angle=270}\\
(a) & (b)
\end{tabular}
\caption{\label{fig:squashed}
The summation indices for the evaluation of the oscillatory part of 
$\theta(0,0|\mbf{\Omega})$ and $\theta(0,0|\hat{\mbf{\Omega}})$ 
with $\epsilon=0.001$. 
In Fig. (a), with $\mbf{\Omega}$, there are 109 summation indices, all
inside a very eccentric ellipse lying along the line $n_2=-n_1$. Figure (b)
corresponds to $\hat{\mbf{\Omega}}$, which was obtained from $\mbf{\Omega}$ 
by way
of a modular transformation \rf{modular}. 
Now only one summation index lies inside the
ellipse: $(n_1,n_2)=(0,0)$. }
\end{figure}

\no This is the example from Appendix C of \cite{dfs}, adapted to the
definition of the theta function used here. It was used in \cite{dfs} to
illustrate the need for a fundamental region of Riemann matrices. The problem
arising is that the ellipsoid (ellipse, in this case) determining the summation
indices in the approximation of the oscillatory part of the Riemann theta
function is very eccentric. The eigenvalues of $\Y$ are 31.0587 and 0.000324,
resulting in an  eccentricity of the ellipse of $1-0.54~10^{-10}$. Thus very
few of the summation indices closest to $(0,0)$ play a part in the evaluation
of the Riemann theta function. This was the problem addressed in \cite{dfs}.
Since our algorithm incorporates a way to determine which summation indices lie
inside the ellipsoid determined by $\Y$, this is not troublesome here. What is
troublesome is that the ellipsoid contains many integer points: already for
$\epsilon=0.001$, the evaluation of $\theta(0,0|\mbf{\Omega})$ requires the
inclusion of 109 terms. This is largely due to the fact that the ellipsoid lies
along a rational direction in the $(n_1,n_2)$ plane, as illustrated in Fig.
\ref{fig:squashed}. 

In such cases, the transformation properties of the Riemann theta function can
be used to great advantage. Just as the elliptic functions and the Jacobian
theta functions, the Riemann theta function has a modular transformation
property. The modular transformation is a transformation on
the Riemann matrix. The transformed Riemann matrix defines a Riemann theta
function, just as the original Riemann matrix does. These two Riemann theta
functions are related: up to an affine transformation of the argument $\mbf{z}$
and an overall scaling factor, they are identical. For details, see
\cite{tata1}. 

\subsection*{Remarks}

\begin{itemize}

\item Geometrically, the modular transformation on the Riemann matrix is a
transformation on the period lattice of the Riemann theta function. Thus the
modular transformation property relates two Riemann theta functions with
different period lattices \cite{tata1}. 

\item For Riemann matrices originating from Riemann surfaces, the modular
transformation amounts to choosing a different canonical intersection basis for
the homology of the Riemann surface \cite{dubrovin}. Then it is not a surprise
that the solutions of differential equations written in terms of Abelian
functions do not depend on this choice. One could think of this as a discrete
symmetry of the solutions of the differential equation. Thus for such
applications it is unimportant to transform the Riemann theta function, as
long as one consistently does all calculations with a preferred homology basis. 

\end{itemize}

Let 
$$
\Gamma=\left(\ba{cc}\mbf{a} & \mbf{b}\\\mbf{c} & \mbf{d}\ea\right)
$$ 
be a symplectic matrix with integer elements, $\Gamma \in SP(2g,\mathbb{Z})$, 
$i.e.$, 
$$
\left(\ba{cc}\mbf{a} & \mbf{b}\\\mbf{c} & \mbf{d}\ea\right)
\left(\ba{cc}\mbf{0}_g & \mbf{I}_g\\-\mbf{I}_g & \mbf{0}_g\ea\right)
\left(\ba{cc}\mbf{a}^T & \mbf{c}^T\\\mbf{b}^T & \mbf{d}^T\ea\right)=
\left(\ba{cc}\mbf{0}_g & \mbf{I}_g\\-\mbf{I}_g & \mbf{0}_g\ea\right),
$$
where $\mbf{a}$, $\mbf{b}$, $\mbf{c}$, and $\mbf{d}$ are $g\times g$ matrices
with integer elements, $\mbf{I}_g$ and $\mbf{0}_g$ are the $g\times g$ identity
and nul matrix, respectively. $SP(2g,\mathbb{Z})$ is called the modular group.
An element of the modular group transforms the Riemann matrix $\mbf{\Omega}$
according to
\beq\la{modular}
\mbf{\Omega}\rightarrow \hat{\mbf{\Omega}}=(\mbf{a}\mbf{\Omega}+\mbf{b})
(\mbf{c}\mbf{\Omega}+\mbf{d})^{-1}.
\eeq
The modular group is generated by the three generators $\Gamma_1, \Gamma_2$ and
$\Gamma_3$ \cite{tata1}:
\bea
\Gamma_1=\left(\ba{cc}\mbf{a} & \mbf{0}_g\\ \mbf{0}_g &
\left(\mbf{a}^{-1}\right)^T\ea\right),\!\! &
\hat{\mbf{\Omega}}=\mbf{a\Omega a}^{-1},&\!\!
\theta(\mbf{A}\mbf{z}|\mbf{A\Omega
A}^T)=\theta(\mbf{z}|\mbf{\Omega}),\\\la{modularadd}
\Gamma_2=\left(\ba{cc}\mbf{1}_g &
\mbf{b}\\ \mbf{0}_g & \mbf{1}_g \ea\right),&\!\! 
\hat{\mbf{\Omega}}=\mbf{\Omega}+\mbf{b},
&\!\!
\theta(\mbf{z}|\mbf{\Omega+b})=
\theta(\mbf{z}+\mbox{diag}(\mbf{b})/2|\mbf{\Omega}),\\
\Gamma_3=\left(\ba{cc}\mbf{0}_g & 
-\mbf{1}_g\\ \mbf{1}_g & \mbf{0}_g \ea\right),&\!\! 
\hat{\mbf{\Omega}}=-\mbf{\Omega}^{-1},
&\!\!
\theta(\mbf{\Omega}^{-1}\mbf{z}|-\mbf{\Omega}^{-1})=\sqrt{\mbox{det}(-i
\mbf{\Omega})}\, e^{\pi i \mbf{z}\cdot\mbf{\Omega}^{-1} \cdot\mbf{z}}
\theta(\mbf{z}|\mbf{\Omega}),
\eea
where diag$(\mbf{b})$ denotes the diagonal part of the matrix $\mbf{b}$, which has
even diagonal elements, and 
$\mbf{b}^T=\mbf{b}$. In the last equation, the branch of the square root is used
which is positive when its argument is positive. 

Since any modular transformation is a composition of these
three types, the transformation formulas on the Riemann theta function 
on the right can be used to calculate the effect of the resulting modular
transformation on the Riemann theta function. 

From our point of view, the goal is to find a transformation on the Riemann
matrix ($i.e.$, a composition of generating modular transformations) to
minimize the eccentricity of  the ellipsoid determined by the transformed
matrix. The algorithm described in this section is due to Siegel
\cite{Siegel3}. Siegel's goal was to construct a fundamental region for Riemann
matrices, analogous to the elliptic case. The algorithm iteratively finds a new
Riemann matrix with improved ($i.e.$, smaller) eccentricity for the ellipsoid
it determines. However, the algorithm is not optimal. An optimal algorithm does
not appear to be known. The description below is an algorithmic description of
Siegel's \cite{Siegel3}, aimed at implementation. This implementation is
included in Maple 8. 

\begin{theo}[Siegel Reduction]
Every Riemann matrix $\mbf{\Omega}$ can, by means of a modular transformation
\rf{modular}, be reduced to a Riemann matrix $\hat{\mbf{\Omega}}=
\hat{\X}+i\hat{\Y}, \hat{\Y}=\T^T \T$, with $\hat{\X}$ ($\hat{\Y}$) the real
(imaginary) part of $\hat{\mbf{\Omega}}$, and
$\T$ upper triangular, such that 
\begin{enumerate}
\item $|\hat{X}_{jk}|\leq 1/2$,$~~~~j,k=1\ldots g$,
\item the length of the shortest lattice vector $\rho$ of the lattice generated by the
columns of $\T$ is bound from below by $\sqrt{\sqrt{3}/2}$, and
\item max$\{|N_j|\,: \|\T\mbf{N}\|\leq R, R>0, \mbox{fixed}, \mbf{N}\in
\mathbb{Z}^g\}$ has an upper bound which only depends on $g$ and $R$. Thus this
upper bound is independent of $\T$.  
\end{enumerate}
\end{theo}

\no The second condition eliminates ellipsoids with high eccentricity. The theorem
in effect guarantees an upper bound for the number of summation indices
required for the evaluation of the oscillatory part of a genus $g$ Riemann
theta function with a prescribed error $\epsilon$ (and thus $R$). This upper
bound does not depend on the reduced Riemann matrix $\hat{\mbf{\Omega}}$. 

\vspace*{0.15in}
\no {\bf Proof:} 
\begin{itemize}
\item {\bf Proof of statement 1:} Using \rf{modularadd}, 
\beq\la{step1}
\mbf{\Omega}\rightarrow \mbf{\Omega}-[\mbox{Re}(\mbf{\Omega})].
\eeq
\item {\bf Proof of statement 2:} We can assume that $\T$ is lattice reduced,
using the LLL algorithm \cite{lll}. Then $\rho=T_{11}=\sqrt{\hat{Y}_{11}}$, since $\T$ is
upper triangular. Siegel \cite{Siegel3} shows that the determinants of the
imaginary parts of two Riemann matrices connected by a modular transformation
\rf{modular} are related by 
\beq\la{siegeldude}
|\mbox{det}(\hat{\Y})|=
\frac{|\mbox{det}({\Y})|}{|\mbox{det}(\mbf{c\Omega+d})|^2},
\eeq
where $\Y$, $\hat{\Y}$ are the imaginary parts of $\mbf{\Omega}$ and 
$\hat{\mbf{\Omega}}$, respectively. Using the modular transformation with
$$
\mbf{a}=\left(\ba{cc}0 & \mbf{0}^T\\ \mbf{0} &\mbf{1}_{g-1}\ea\right),
\mbf{b}=\left(\ba{cc}-1 & \mbf{0}^T\\ \mbf{0} &\mbf{0}_{g-1}\ea\right),
\mbf{c}=\left(\ba{cc}1 & \mbf{0}^T\\ \mbf{0} &\mbf{0}_{g-1}\ea\right),
\mbf{d}=\left(\ba{cc}0 & \mbf{0}^T\\ \mbf{0} &\mbf{1}_{g-1}\ea\right),
$$
with $\mbf{0}$ the $(g-1)$-dimensional zero vector, \rf{siegeldude} becomes 
$$
|\mbox{det}(\hat{\Y})|=\frac{|\mbox{det}({\Y})|}{|\Omega_{11}|^2}.
$$
If $|\Omega_{11}|<1$, this transformation is applied. This transformation
preceeded by the transformation giving \rf{step1} is repeated until
$|\hat{\Omega}_{11}|\geq 1$. Then
$|\hat{\Omega}_{11}|^2=\hat{X}_{11}^2+\hat{Y_{11}}^2\geq 1$. Thus
$\rho=\sqrt{\hat{Y}_{11}}\geq\sqrt{\sqrt{1-\hat{X}_{11}^2}}\geq \sqrt{\sqrt{3}/2}$,
since $\hat{X}_{11}^2\leq 1/4$. It remains to be shown that this iteration
terminates. This is demonstrated by Siegel \cite{Siegel3}. 
\item {\bf Proof of statement 3:} If $\T$ is lattice reduced using the LLL
algorithm, then (see \cite{lll})
\begin{itemize}
\item[$\ast$] $T_{jj}>0$, $~~~~~~~~~~~~~~~~~~~~~~~1\leq j\leq g$, 
\item[$\ast$] $|T_{jk}|\leq |T_{jj}|/2$, $~~~~~~~~~~~~~~1\leq j<k\leq g$, 
\item[$\ast$] $T_{j,j+1}^2+T_{j+1,j+1}^2\geq T_{jj}^2$, $~~~1\leq j < g$.
\end{itemize}
It follows from these last two properties that $T_{jj}\geq
\frac{\sqrt{3}}{2}T_{j-1,j-1}$, for $j\in (1,g]$, and thus $T_{jj}\geq
\left(\frac{\sqrt{3}}{2}\right)^{j-1}\rho$. Then   
max$\{|N_j|\,: \|\T\mbf{N}\|\leq R, R>0, \mbox{fixed}, \mbf{N}\in
\mathbb{Z}^g\}$ $\leq$  max$\{|N_j|\,: |(\T\mbf{N})_j|\leq R, 
R>0, \mbox{fixed}, j\in [1,g], \mbf{N}\in
\mathbb{Z}^g\}$ $\leq$ $R/r$. \hfill $\bbox$
\end{itemize}

\no Note that the preceding proof indeed shows that the new ellipsoid
defined by $\|\T\mbf{N}\|=R$ has an eccentricity bound away from 1. 

This algorithm was used on the Riemann matrix given in \rf{rm2}, with great
success: after applying the algorithm, the evaluation of the oscillatory part
of the Riemann theta function parametrized by the transformed Riemann matrix at
$\mbf{z}=(0,0)$ with absolute error $\epsilon=0.001$ requires only one summation
index: $(n_1,n_2)=(0,0)$. The modular transformation used is 
$$
\Gamma=\left(
\ba{rrrr}
0&0&8&7\\0&0&7&6\\6&-7&0&0\\-7&8&0&0
\ea
\right),
$$
resulting in
$$
\hat{\mbf{\Omega}}=\left(
\ba{cc}
7.94597&-3.94937\\
-3.94937&14.4545
\ea
\right),
$$
giving an ellipsoid with eccentricity 0.927921, with a major-axes ratio 
$\approx 3/8$. 

\section*{Acknowledgements} 

The research presented in this paper was supported through NSF grants
DMS-9805983, DMS-0071568 (BD), SFB288 (Differential geometry and quantum
physics, AB, MH, MS) and DMS-0098034 (MvH).

\section*{Appendix A: The Maple implementation} 

The maple code can be viewed by typing the following commands in Maple 8 or
later:
\begin{verbatim}
interface(verboseproc=2);
op(RiemannTheta);
op(`RiemannTheta/doit`);
op(`RiemannTheta/boundingellipsoid`);
op(`RiemannTheta/findvectors`);
op(`RiemannTheta/make_proc`);
op(`RiemannTheta/finitesum`);
\end{verbatim}
It can also be downloaded from {\tt 
http://www.math.fsu.edu/\verb+~+hoeij/RiemannTheta/}.

The procedure {\tt RiemannTheta} uses a variety of arguments. 
The first two of these are required. All others are optional and can be omitted.
\begin{enumerate}
\item A $g\times g$ Riemann matrix $\mbf{\Omega}$.
\item A $g$-dimensional vector $\mbf{z}$.
\item A number $\epsilon$ indicating the desired
accuracy of the result.
\item A (possibly empty) list of vectors. Each vector is used as in
\rf{dirders} for the calculation of directional derivatives. 
\item Other optional arguments:  The algorithm computes
the exponential growth factor $a$ and
the oscillatory part $b$ separately.
The user can specify if $a$ and $b$ should both be given
in the output, or if the output should consist of only ${\rm exp}(a)b$.
\end{enumerate}

{\tt RiemannTheta} distinguishes two cases:
\begin{itemize}
\item[a.] $z$ evaluates to an element of $\C^g$.
\item[b.] $z$ does not evaluate to an element of $\C^g$, 
because it contains one or more variables that have no assigned values.
\end{itemize}
 
\noindent If no directional derivatives are given, then {\tt RiemannTheta}
computes either a list of two complex numbers $[a,b]$ or one complex number
$e^a b$ such that $\theta(\Omega, z) \approx e^a b$. Here $b$ is
bounded for $z \in \C^g$, and has error less than $\epsilon$. If directional
derivatives are given, then $b$ grows polynomially. For case b, it is not 
guaranteed that $b$'s error is bounded by $\epsilon$ unless an a-priori
bound for $z$ is known.

If {\tt RiemannTheta} is called, then the set $S$ of points in the
``ellipsoid'' described in Figure \ref{fig:ellipsoids} is calculated. See
Figure \ref{fig:ellipsoids}(a) for case a, and Figure \ref{fig:ellipsoids}(b)
for case b. Following the computation, {\tt RiemannTheta} returns an expression
that contains {\tt `RiemannTheta/finitesum`(args)}. Here {\tt
`RiemannTheta/finitesum`} is a procedure and {\tt args} is a list of arguments.
These arguments contain all information necessary to compute the value of
RiemannTheta with the specified accuracy. This information includes $\mbf{z}$,
$\mbf{\Omega}$ and $S$, but reorganized into a form more convenient for
summation ($e.g.$, the real and imaginary parts of the input are separated, as
are the integer and non-integer part before summation over the set $S$ is
begun). However, the summation is not done unless $\mbf{z} \in \C^g$.  Only if
$\mbf{z}$  contains no unassigned variables can $\mbf{z}$ be evaluated to an
element of $\C^g$, and only in this case will {\tt
`RiemannTheta/finitesum`(args)} return a number. This number then becomes the
output of {\tt RiemannTheta}. If $\mbf{z}$ does contain unassigned variables,
then the summation will be delayed, and {\tt `RiemannTheta/finitesum`(args)} 
will remain unevaluated. Thus, using variables in $\mbf{z}$ instead of complex
numbers causes {\tt RiemannTheta} to return an answer that contains an
unevaluated procedure. This is useful if one wants to calculate more than one
value of {\tt RiemannTheta} for the same $\mbf{\Omega}$.  If  only a single
value of {\tt RiemannTheta} is wanted, one gives {\tt RiemannTheta} a vector 
$\mbf{z} \in \C^g$. If one wants to compute {\tt RiemannTheta} for many
$\mbf{z}$'s  ($e.g.$, for plotting purposes) then one includes variables in
$\mbf{z}$. The procedure which is the output of {\tt RiemannTheta} is then used
for evaluation.  This way the set of points in Figure \ref{fig:ellipsoids}(b) 
is computed only once, which is more efficient than computing the points in
Figure \ref{fig:ellipsoids}(a) for each value of $\mbf{z}$.

Below is a verbatim example of interactive use of Maple and {\tt RiemannTheta}. 

\newpage

\begin{verbatim}
> r3 := sqrt(-3)/3:
> M := evalf( Matrix(2,2,[[1+2*r3,-1-r3],[-1-r3,1+2*r3]]) );
                   [ 1. + 1.154700539 I     -1. - 0.5773502693 I]
              M := [                                            ]
                   [-1. - 0.5773502693 I     1. + 1.154700539 I ]
> eps := 0.001:
> z := [1-I,  I+1]; # Here z contains no variables.
                              z := [1 - I, 1 + I]
> R := RiemannTheta(z, M, [], eps);
                                                       -8
                    R := -21.76547256 - 0.2323298326 10   I
> L := RiemannTheta(z, M, [], eps, output=list);
                                                             -10
           L := [3.627598727, -0.5785248137 - 0.6175311504 10    I]
> a := L[1]:
> b := L[2]:
> R := exp(a)*b;
                                                       -8
                    R := -21.76547256 - 0.2323298326 10   I
> z := [X,    I+1];   # Now z contains 1 variable.
                                z := [X, 1 + I]
> R := RiemannTheta(z, M, [], eps, output=list);
                       2
R := [3.627598726 Im(X)  + 3.627598724 Im(X) + 3.627598726, 
     RiemannTheta/finitesum( ... ) ]
> eval(R, X=1-I);
                                                            -10
             [3.627598728, -0.57852736323 - 0.61753198638 10    I]
> eval(R, X=1-2*I);
                                                            -9
              [10.88279618, 0.62464131574 - 0.56009279234 10   I]
> eval(R, X=1-3*I);
                                                           -10
             [25.39319109, 0.44006321314 - 0.92901122197 10    I]
\end{verbatim}

\section*{Appendix B: The Java implementation}

The Java implementation is an adaptation of an earlier  implementation in C
which has been in use since 1994. The Java version uses the definition
\rf{theta} of the Riemann theta function.  The C version employed a different
definition:

$$
\tilde{\theta }\left( \mbf{z}|\mbf{\Omega} \right) 
=\sum _{\mbf{n}\in \mathbb{Z}^{g}}
e^{\frac{1}{2}\mbf{n}\cdot \mbf{B} \cdot \mbf{n}+\mbf{z}\cdot \mbf{n}}
=\theta \left( \frac{\mbf{z}}{2\pi i}\Big|\frac{\mbf{B}}{2\pi i}\right),
$$
where the real part of $\mbf{B}$ is negative definite.

\begin{figure}
\centerline{\psfig{file=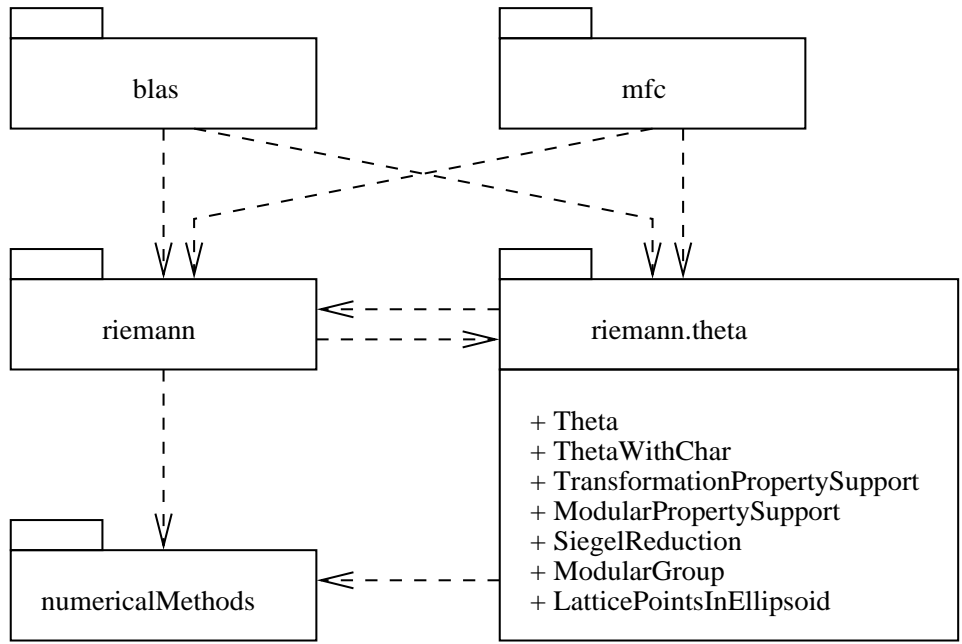,width=5in}}
\caption{\label{fig: dependencies of riemann.theta} 
Dependencies of the package {\tt riemann.theta}.}
\end{figure}

The java implementation of the Riemann  theta function is realized in the
package {\tt riemann.theta}, which includes the public classes {\tt
LatticePointsInEllipsoid}, {\tt ModularGroup}, {\tt SiegelReduction}, {\tt
ModularPropertySupport},  {\tt TransformationPropertySupport}, {\tt Theta} and
{\tt ThetaWithChar}. Unlike other programming languages,  Java does not
incorporate a complex type, thus the package requires an implementation of
such, realized in {\tt mfc.number.Complex}.  The implementation also requires
the packages {\tt blas} (basic linear algebra system), and {\tt
numericalMethods}. This last package is used only in internal computations. 
Figure \ref{fig: dependencies of riemann.theta} illustrates these dependencies,
showing only the classes within {\tt riemann.theta}.

The class {\tt Theta} implements the uniform approximation theorems
\ref{theo:uniform}, \ref{theo:unifder} and \ref{theo:unifderder}. The default
error used by the program is the FFE, with default error tolerance set to
$10^{-7}$.  Siegel's reduction algorithm as described in section
\ref{sec:siegel}  is always used. 

\vspace*{0.2in}

{\bf Example:} Consider the slices of a genus 2 Riemann theta function shown 
in Figure \ref{fig:plots}. The complete Java program for computing the data in
one of these slices is: 

{\tt 
\medskip{}

\hskip  2mm import riemann.theta.Theta;

\hskip  2mm import mfc.number.Complex;

\hskip  2mm import blas.ComplexVector;

\hskip  2mm import blas.ComplexMatrix;

\hskip  2mm public class ExampleSlice \{

\hskip  2mm public static void main( String {[}{]} argv ) \{ 
 
\hskip  7mm {// create the period matrix}

\hskip  7mm ComplexMatrix B = new ComplexMatrix( 2 ); 

\hskip  7mm {// set the period matrix}

\hskip  7mm B.set( 0,0, 1.690983006, 0.9510565162 );

\hskip  7mm B.set( 0,1, 1.500000000, 0.3632712640 );

\hskip  7mm B.set( 1,0, 1.500000000, 0.3632712640 );

\hskip  7mm B.set( 1,1, 1.309016994, 0.9510565162 );

\hskip  7mm // use the different normalization

\hskip  7mm B.setTimes( new Complex( 0, 2 {*} Math.PI ) );

\hskip  7mm {// create a complex argument vector}

\hskip  7mm ComplexVector V = new ComplexVector( 2 );

\hskip  7mm {// create a theta function instance with period matrix }

\hskip  7mm Theta theta = new Theta( B ); 

\hskip  7mm {// create storage for the slice}

\hskip  7mm Complex{[}{]}{[}{]} slice = new Complex{[}101{]}{[}101{]}; 

\hskip  7mm {// loop over grid}

\hskip  7mm for( int i=0; i<101; i++ )

\hskip 12mm for( int j=0; j<101; j++ ) \{

\hskip 17mm {// set grid vector}

\hskip 17mm V.set( 1, 2 {*} i {*}Math.PI / 100, 2 {*} j {*} Math.PI / 100 ); 

\hskip 17mm {// evaluate theta function at grid vector}

\hskip 17mm slice{[}i{]}{[}j{]} = theta.theta( V ); 

\hskip 12mm \}

\hskip  7mm \}

\hskip  2mm \}

\medskip{}
}

\no To compile this example, the user needs the Java archives {\tt riemann.jar},
{\tt blas.jar}, {\tt mfc.jar} and {\tt numericalMethods.jar} in their
classpath. These archives, as well as their source code and documentation
is available from {\tt www-sfb288.math.tu-berlin.de/\verb+~+jem}.

%
%
%
%
%

The package {\tt riemann.theta} also includes a class {\tt ThetaWithChar} which
incorporates Riemann theta functions with characteristics. 

\bibliographystyle{plain}

\newpage

\no Alexander Bobenko,
Matthias Heil,
Markus Schmies\newline
Fachbereich Mathematik\newline
Technische Universit\"at Berlin\newline
Strass des 17.Juni 136\newline
10623 Berlin, Germany\newline
bobenko@math.tu-berlin.de\newline
matt@heil-lanzinger.de\newline
schmies@sfb288.math.tu-berlin.de

\vspace*{1cm}

\no Bernard Deconinck\newline
Department of Mathematics\newline
Colorado State University\newline
Fort Collins, CO 80524-1874, USA\newline
deconinc@math.colostate.edu

\vspace*{1cm}

\no Mark van Hoeij\newline
Department of Mathematics\newline
Florida State University\newline
Tallahassee, FL 32306, USA\newline
hoeij@math.fsu.edu
\end{document}